\documentclass[preprint]{aa} 

\usepackage[varg]{txfonts} 
\usepackage{natbib}

\usepackage{amsmath}
\usepackage{amstext}
\usepackage{graphicx}
\usepackage{hyperref}

\newcommand{\nuc}[2]{\ensuremath{^{#1}}#2}
\newcommand{\gcm}[0]{g\ensuremath{/}cm\ensuremath{^3}}
\newcommand{\Nepa}[0]{\ensuremath{^{22}}Ne\ensuremath{+ \alpha}}
\newcommand{\SSC}[0]{S\ensuremath{_{\text{scale}}}}
\newcommand{\YTMIN}[0]{\ensuremath{y_{\text{t,min}}}}
\newcommand{\EPS}[0]{\ensuremath{\epsilon}}
\newcommand{\YSCALE}[0]{\ensuremath{y_{\text{scale}}}}

\begin{document}

\title{Performance Improvements for Nuclear Reaction Network
  Integration}
\author{R. Longland\inst{1,2}\thanks{\email{richard.longland@upc.edu}}
  \and D. Martin\inst{1,2} \and J. Jos\'e\inst{1,2}}
       
\institute{
 Departament de F\'{\i}sica i Enginyeria Nuclear, EUETIB,
  Universitat Polit\`{e}cnica de Catalunya, c/~Comte d'Urgell 187,
  E-08036 Barcelona, Spain\label{inst1} \and
  Institut d'Estudis Espacials de Catalunya (IEEC), Ed. Nexus-201,
  C/~Gran Capit\`{a} 2-4, E-08034
  Barcelona, Spain\label{inst2}}

\abstract{} 
{  The aim of this work is to compare the performance of three reaction
  network integration methods used in stellar nucleosynthesis
  calculations. These are the Gear's backward differentiation method,
  Wagoner's method (a 2nd-order Runge-Kutta method), and the
  Bader-Deuflehard semi-implicit multi-step method.}
{To investigate the efficiency of each of the integration methods
  considered here, a test suite of temperature and density versus time
  profiles is used. This suite provides a range of situations ranging
  from constant temperature and density to the dramatically varying
  conditions present in white dwarf mergers, novae, and x-ray
  bursts. Some of these profiles are obtained separately from full
  hydrodynamic calculations.  The integration efficiencies are
  investigated with respect to input parameters that constrain the
  desired accuracy and precision.}
{Gear's backward differentiation method is found to improve
  accuracy, performance, and stability in integrating nuclear reaction
  networks. For temperature-density profiles that vary strongly with
  time, it is found to outperform the Bader-Deuflehard method
  (although that method is very powerful for more smoothly varying
  profiles). Wagoner's method, while relatively fast for many
  scenarios, exhibits hard-to-predict inaccuracies for some choices of
  integration parameters owing to its lack of error estimations.}
 {}

\keywords{ Methods: numerical - Nuclear reactions, nucleosynthesis,
  abundances}

\maketitle

\section{Introduction}
\label{sec:intro}

Stellar models are an integral part of astrophysics that help us
understand the structure and evolution of stars and their
nucleosynthesis output. In order to properly resolve the wide range of
conditions within these environments, models must be computed with
high resolution in both space and
time. Stellar models rely on a number of conservation equations (i.e., mass,
energy, momentum) and energy transport mechanisms (i.e., radiation,
convection, and conduction), which depend on the physical condition of
the star.  Further complicating these models, energy generation
through nuclear reactions must be taken into account. Depending on the
number of isotopes and nuclear interactions involved, these reactions
can become very computationally expensive. Time evolution of
hydrodynamical models is sometimes restricted, therefore, by the
solution of a large system of ordinary differential equations that
describe the network of nuclei in the star. If detailed nuclear
reaction networks are desired, hydrodynamic models of the stellar
system under investigation can be slowed to the point that they are no
longer viable without high power supercomputers, unless
post-processing techniques are employed.

The network of nuclei and reactions required in nucleosynthesis
investigations can be divided into two broad nuclear reaction
networks: (i) that which is responsible for energy generation and is
therefore essential for any accurate hydrodynamical evolution of the
system. Examples of these networks were presented by
\cite{WEA78,MUE86,TIM00}; and (ii) a reaction network that is only
necessary for detailed computation of nucleosynthesis. This second
network then uses, as input physics, the hydrodynamical properties of
a plasma (i.e., temperature and density) obtained from a separate code
using a limited set of reactions. De-coupling the networks in this way
allows us to compute the expensive hydrodynamical model only
once\footnote{Other techniques, such as parallelisation of
  hydrodynamic models can be of benefit in improving model computation
  time~\citep{MAR13}.}, the output of which can be used to perform
many post-processing calculations. The advantages of this approach are
especially obvious for nucleosynthesis sensitivity studies \citep[see,
for example,][]{HIX03,ROB06,PAR08}.  This method of de-coupling the
nuclear reaction networks assumes that any nucleosynthesis occurring
in the second step does not affect the behaviour of the hydrodynamical
models in the first.  In some cases, this decoupling of hydrodynamical
models from nucleosynthesis networks is not valid. For example, the
1-D implicit hydrodynamic code ``SHIVA''~\citep{JOS99} considered here
integrates a network of over 1400 nuclear processes and several
hundred isotopes into models of x-ray bursts, making nucleosynthesis a
considerable bottleneck in the computational efforts. With these
restrictions in mind, improving the reaction network integration
performance is desirable in either of the cases described above,
whether it is allowing larger step-sizes in full hydrodynamical models
or increasing post-processing performance. Since full 1-D
  hydrodynamic models are time-consuming to compute owing to their
  high spacial resolution, we concentrate our efforts on
  post-processing calculations using temperature-density profiles
  obtained separately.  This simplification allows us to perform a
  detailed investigation of the sensitivity of the integration method
  to parameters controlling accuracy and precision. Furthermore, it
  allows us to investigate the performance of the integration method
  for computing models of a wide range of astrophysical
  environments\footnote{However, any performance increases found in
  this work may also yield gains in computation time in full
  hydrodynamic models depending on the details of the code in
  question. This will be discussed in detail in a future paper.}.

Nuclear reaction networks are particularly difficult to integrate
numerically because the ordinary differential equations governing
their evolution are ``stiff''. A set of equations
is considered stiff if the solution depends strongly on small
variations in some of the terms. Consequently, very small step-sizes
are required to ensure solution stability when explicit methods are
used\footnote{For a recent and detailed discussion of improving the
  performance and stability of explicit methods, the reader is
  referred to \cite{GUI12}.}. For complex nuclear networks, these
step-sizes can become unacceptably small. Implicit methods are usually
used to alleviate these problems~\citep{NumericalRecipes}. These
difficulties faced in the numerical integration of nuclear reaction
networks were identified and successfully conquered in the late 1960's
using a range of semi-implicit and implicit methods
\citep{TRU67,ARN69,WAG69,WOO73}. While these methods allowed nuclear
reaction network evolution to be performed, the techniques involved
are rather primitive, and possible improvements have been largely
ignored for 40 years.

While most nucleosynthesis reaction network codes still employ the
methods mentioned above, some groundbreaking work has been done on
investigating improved techniques. \cite{TIM99} investigated a large
number of linear algebra packages and semi-implicit time integration
methods for solving nuclear reaction networks. The findings of that
investigation were important not only in improving network integration
efficiency, but also in characterising the accuracy of abundance
yields. This second advantage is a key point in constructing more
robust and flexible integration algorithms. One integration method
that \cite{TIM99} did not investigate, however, was Gear's backward
differentiation method~\citep{GEA71}, which is a fully implicit method
that utilises past history of the system of ordinary differential
equations to predict future solutions, thus increasing the integration
time-step that can be accurately taken.

It is the purpose of this paper to investigate a number of methods for
integrating nuclear reaction networks in stellar codes. We will
concentrate on 3 methods, namely Wagoner's two-step
method~\citep{WAG69}, Gear's backward differentiation method\
\citep{GEA71}, and the Bader-Deuflehard
method~\citep{BAD83}. Wagoner's method will be considered primarily
because it is widely used in nuclear astrophysical codes, hence
serving as a convenient baseline for comparison.  Gear's method has
been utilised previously in a number of studies: in big bang
nucleosynthesis~\citep{ORL00}; explosive
nucleosynthesis~\citep{LUM89}; \nuc{26}{Al} production~\cite{WAR80};
and r-process calculations~\citep{NAD98,PAV09}, but it's performance
for a range of nucleosynthesis applications has never been fully
investigated. Finally, the Bader-Deuflehard method was suggested
by~\cite{TIM99}, which was found to dramatically improve integration
times for the profile considered in that work. Indeed, this method has
been successfully used in a number of
studies~\citep[e.g.,][]{NOE07,STA12,PAK12}.

Each of the integration methods will be introduced in
Sec.~\ref{sec:methods}, for which we will highlight the important
detail of step-size adjustment and error checking. The test suite of
profiles and reaction networks will be introduced in
Sec.~\ref{sec:tests}, and the results of these tests presented in
Sec.~\ref{sec:result}. Detailed discussion of the results is given in
Sec.~\ref{sec:discussion}, paying attention to numerical reasons for
any behavioural differences between the methods. Conclusions and
recommendations are made in Sec.~\ref{sec:conclusions}.

\section{Integration Methods}
\label{sec:methods}

A large number of methods can be used for implicitly or
semi-implicitly solving stiff systems of ordinary differential
equations \citep[see, for example, ][]{NumericalRecipes}. In this
work, we will concentrate on three of these possibilities: (i) Wagoner's two-step integration technique (basically, a second
order Runge-Kutta method), which is commonly known in numerical
astrophysics~\citep{WAG69}. This method hinges on taking a large
number of computationally inexpensive time-steps to solve the system
of equations; (ii) the Bader-Deuflehard method, which is a
semi-implicit method that is based on taking computationally expensive
time-steps, but offset by only requiring few of them; and (iii) Gear's
backward differentiation method, which uses previous history of the
system of equations to predict future behaviour, thus allowing larger
steps to be taken while preserving the fairly inexpensive individual
steps exhibited by Wagoner's method. Note that this final
  method is only applicable in cases where past history can be stored
  in this way.

For each of the integration methods discussed, the most
computationally expensive procedure for reaction network integration
is solving the system of equations that describes the rate of
change of each nucleus as a function of every other nucleus. To
perform this task, \cite{TIM99} recommended the linear
algebra package, \texttt{MA28}. Since that time, an updated version to
the package, \texttt{MA48}~\citep{MA48}, has become available, which
we use in the current investigation. Computational efficiency
differences between the two are expected to be minor, with the main
advantages of the new package being ease-of-use.

\subsection{Wagoner's Method}
\label{sec:th-wagoner}

What is often referred to as ``Wagoner's method'' in nuclear
astrophysics is outlined by \cite{WAG69}. The method is a
semi-implicit, second-order Runge-Kutta method, and represents an
extension of the single-order method first used in~\cite{TRU67}
and~\cite{ARN69}. Below, we present an outline of the method to
clarify notation, allowing easier comparison with the other methods
discussed below.

To construct the system of ordinary differential equations, we first
define a vector, $y_n$, which represents the molar fraction of all
nuclei in the network at time $t_n$, i.e.,
$y_n=[Y_1,Y_2,\ldots,Y_{i,max}]_n$. Here, the molar fraction of an
isotope is defined as $Y_i=X_i/A_i$, where $X_i$ is the mass fraction
and $A_i$ is the atomic mass of species $i$. These abundances are
advanced to time $t_{n+1}$ with a time-step of $h$ by solving the
equation:
\begin{equation}
  \label{eq:wag1}
  y_{n+1} = y_n + \frac{1}{2} h \left[ \frac{dy}{dt}(y_n,t_n) +
           \frac{dy}{dt}(\tilde{y}_{n+1},t_{n+1}) \right],
\end{equation}
where $\tilde{y}_{n+1}$ are the intermediate abundances evaluated at
the future time, $t+1$. The time derivatives of $y$ are found by
considering all reactions that produce and destroy each nucleus. These
can be split into three contributions: (i) decay processes involving
single species, (ii) two-body reactions of the form $A(a,b)B$, and
(iii) three body reactions such as the triple-alpha process. Putting
these together:
\begin{eqnarray}
  \label{eq:dydt}
  \frac{dY_i}{dt} &=& \sum_j N_i \lambda_j Y_j + \sum_{j,k}
  \frac{N_i}{N_j! N_k!} \rho N_A \langle \sigma v \rangle_{j,k} Y_j
  Y_k + \nonumber \\
  & & \qquad \sum_{j,k,l} \frac{N_i}{N_j! N_k! N_l!} \rho^2 N_A^2 \langle
  \sigma v \rangle_{j,k,l} Y_j Y_k Y_l 
\end{eqnarray}
Here $N_i$ is an absolute number denoting the number of the species
$i$ that is produced in the reaction. Note that this can be a negative
number for destructive processes. The decay rate for species $i$ is
given by $\lambda_i$, $\rho$ denotes the density of the environment,
$N_A$ is Avogadro's number, and $\langle \sigma v \rangle_{i,j}$ is
the reaction rate per particle pair for two-body reactions involving
species $i$ and $j$. Three-body reaction rates follow the same
notation see \citep[see][for a more detailed discussion]{ILIBook}.

Equation~(\ref{eq:wag1}) is solved using a two-step procedure in which
the second part of the equation depends on the first part through
$\tilde{y}_{n+1}$:
\begin{equation}
  \label{eq:wag2}
  \tilde{y}_{n+1} = y_n + h \frac{dy}{dt}(y_n,t_n)
\end{equation}
To ameliorate numerical instabilities arising from solving these
equations explicitly, these equations are solved implicitly. To
achieve this, a Jacobian matrix must first be constructed. If
$f(y_n,t_n) \equiv dy_n/dt$, then the Jacobian matrix is $\mathbf{J}
= \partial f/ \partial y$. Using this, Eq.~(\ref{eq:wag2}) can be
re-written using an implicit scheme and represented in matrix form by
\begin{equation}
  \label{eq:wag3}
  \left[\mathbf{I} - h\mathbf{J}\right]\tilde{y}_{n+1} = y_n,
\end{equation}
where $\mathbf{I}$ is the unity matrix. This equation is subsequently
solved to obtain the new abundances, $\tilde{y}_{n+1}$ and
Eq.~(\ref{eq:wag1}) is solved by performing this operation
twice. Solving Eq.~(\ref{eq:wag3}) represents the most
computationally expensive operation in each of the methods considered
here. This large $M\times M$ matrix (where $M$ is the number of nuclei
in the network) exhibits a characteristic sparse
pattern. While, in principle, every nucleus in the network can
  interact with every other, processes involving light particles are
  far more likely owing to their smaller Coulomb barrier. Large
  portions of the matrix, therefore, are essentially negligible.
Typical dimensions of this matrix are, for example, $\sim 100 \times
100$ for novae, or $\sim 600 \times 600$ for x-ray bursts. 
These Jacobian patterns are
discussed in considerable detail by \cite{TIM99}. The solution of
these sparse systems is left to the well optimised \texttt{MA48}
routines from the HSL Mathematical Software Library~\citep{MA48} in
the present work. Since the solution of these large systems of
  equations is by far the most expensive operation in integrating
nuclear reaction networks, reducing the total number of times this
operation must be performed is the ultimate goal when attempting to
increase algorithm efficiency. However, we must develop ways to ensure
that safe step-sizes are used to obtain accurate results.

Safe step-sizes in Wagoner's method are usually computed in an ad-hoc
way based on the largest abundance changes in the previous step. For
example, the new step-size, $h'$, can be computed using
\begin{equation}
  \label{eq:newstep}
  h' = K h \left[\frac{y_{n+1}}{y_{n+1}-y_{n}}\right]_{\text{min}},
\end{equation}
where $K$ is varied by hand to ensure convergence of the final
abundances. It typically assumes values between 0.1 and 0.4 (for
the test suites in Sec.~\ref{sec:tests}, a value of $K=0.25$ is
adopted). To avoid small abundances dominating the step-size
calculations, a parameter, $y_{\text{t,min}}$, is introduced. Only
nuclei with abundances greater than this value are used to calculate
step-sizes, with others being allowed to vary freely. Furthermore, a
parameter can be introduced to limit the maximum step-size,
represented in our implementation by the parameter \SSC, which
controls the minimum number of time-steps allowed. This parameter
ensures that any sudden changes in the profile are not passed
unnoticed.

While Wagoner's method is slightly more advanced than a simple Euler
method \citep[see][]{TIM99}, it still provides no good means of
determining step accuracy. In order to ensure accurate integration, a
combination of experience and conservatism is required, in which
abundance changes and the rate of step-size increases are severely
restricted. The only available method of checking solution accuracy is
by considering the conservation of mass in the system. Once the total
mass deviates outside a given tolerance (imposed by checking that
$\sum X_i = 1$), integration is assumed to have failed, and a new
calculation with more restrictive integration parameters must be
attempted.

\subsection{Bader-Deuflehard Semi-implicit Method}
\label{sec:th-bader}

The Bader-Deuflehard method \citep{BAD83,TIM99,NumericalRecipes}
relies on the semi-implicit mid-point rule:
\begin{equation}
  \label{eq:midpoint-rule}
  \left[1-h\frac{\partial f}{\partial y}\right] \cdot y_{n+1} = 
  \left[1-h\frac{\partial f}{\partial y}\right] \cdot y_{n-1} + 
  2 h \left[ f(y_n) - \frac{\partial f}{\partial y}\cdot y_n \right]
\end{equation}
where $f(y_n)$ is the time derivative of the isotopic abundance
vector, $y_n$, at time $t_n$. Rather than solving this equation once
per time-step, the adopted strategy is to solve several steps at once.
Consequently, one large step of length $H$ can be taken in $m$
sub-steps, each of length $h$. The method relies on the basic
assumption that the solution following a step $H$ is a function of
the number of sub-steps, which can be probed by solving the equations
for a range of trial values for $m$. Once this function has been
found, the solution can be extrapolated to an infinite number of
sub-steps, thus yielding converged abundances. \cite{BAD83} developed
the sequence of $m$ values that provides best convergence, so the
number of sub-steps is varied in the range:
$m=2,6,10,14,22,34,50$. After each attempt, the result is extrapolated
to an infinite number of sub-steps until accurate convergence (within
some pre-defined tolerance) is obtained. For a specific value of
$m$, the solution after a large step $H$ is given by
$y_{n+1}=y_n+\sum_m \Delta_m$, the sub-steps to be taken are:
\begin{eqnarray}
  \label{eq:BaderStepping}
  k=0 &\,& \Delta_0 = \left[1-h\frac{\partial f}{\partial y}\right]^{-1}
  \!\!\!\!\cdot h f(y_0) \\
  k=1 \ldots m-1 &\,& \Delta_k = \Delta_{k-1} + 2 \left[1-h\frac{\partial f}{\partial y}\right]^{-1} \!\!\!\!\cdot h (f(y_k) - \Delta_{k-1}) \\
  k=m &\,&  \Delta_m = \left[1-h\frac{\partial f}{\partial
      y}\right]^{-1} \!\!\!\!\cdot h (f(y_m) - \Delta_{m-1})
\end{eqnarray}
If no convergence is reached with $m=50$, the step is re-attempted
with a smaller step-size, $H$. This takes into account cases in which
local environment changes occur within the step (i.e., there is a
sharp temperature change in the stellar environment). In these cases,
the assumption that the extrapolated solution is a function of $m$ is
no longer valid, so smaller steps must be attempted until convergence
is obtained.

Obviously, this method requires that the system of equations
  is solved a large number of times, and even for the best case
scenario, convergence is reached when $m=6$. In the worst case, when
$m=50$, a total of 138 Lower Upper (LU) matrix decompositions
must be performed to take a single step. This method, therefore,
relies on large steps to offset their computational cost
\citep[see][]{TIM99}.

Error estimation and step-size adjustment for the Bader-Deuflehard
method must also be considered. The error estimate, $\epsilon_k$, at
the current sub-step order $k$, is estimated directly by the
polynomial extrapolation truncation error~\citep[see chapter 3
in][]{NumericalRecipes}. Consequently, if the desired accuracy is
$\epsilon$, the new step-size, $H_k$, for this order can be estimated
by
\begin{equation}
  \label{eq:bader-newstepsize}
  H_k = H \left(\frac{\epsilon}{\epsilon_k}\right)^{1/2k+1}
\end{equation}
However, we must consider the work required to reach convergence for
different values of $k$. The number of function evaluations (expected
to dominate the computational cost of the method) required to compute
a step is $A_{k+1}$. The work required per unit time-step of order,
$k$, is therefore
\begin{equation}
  \label{eq:bader-work}
  W_k = \frac{A_{k+1}}{H_k} H,
\end{equation}
where the current step-size $H$ is used to normalise the
calculation. The order used to estimate the next step-size is chosen
by identifying the value of $k'$ that minimises $W_k$. The new
time-step is then computed using this value for $H_{k'}$ in
Eq.~(\ref{eq:bader-newstepsize}).

When performing the above calculations, it is customary to scale the
calculated errors for each nucleus by their abundance to obtain
relative errors, $\bar{\epsilon_i}=\epsilon_i/Y_i$, to be used in
Eq.~(\ref{eq:bader-newstepsize}). However, we do not wish to weight
nuclei with very small abundances equally in these step-size
calculations, so an additional parameter, $y_{\text{scale}}$, is
introduced. For abundances below this value, relative errors are
normalised to $y_{\text{scale}}$ such that:
\begin{equation}
  \label{eq:yscale}
  \bar{\epsilon}_i =
  \begin{cases}
    \epsilon_i/Y_i & \text{if } Y_i > y_{\text{scale}}\\
    \epsilon_i/y_{\text{scale}} & \text{if } Y_i \le y_{\text{scale}}
  \end{cases}
\end{equation}

\subsection{Gear's Backward Differentiation Method}
\label{sec:th-gear}

The backward differentiation method developed by \cite{GEA71} hinges
on using past behaviour of the system to predict the solution at
future times. Following that prediction, Newton-Raphson iteration is
used to correct the solution until a predefined precision is
reached. The advantage of this method is that it allows an increase in
possible step-size, while maintaining a limited number of matrix
LU-decompositions. For a good discussion of the method, including the
particular implementation of Gear's method adopted in the present
paper, the reader is referred to \cite{BYR75}. Here, we discuss the
key characteristics of the method.

Gear's method hinges on updating and storing a q$^{\text{th}}$ order
vector of past behaviour known as the Nordsieck vector~\citep{NOR61},
$z_n$, defined at each time $t_n$ as
\begin{equation}
  \label{eq:nord}
  z_n=\left[y_n, h\dot{y}_n, h^2\ddot{y}_n, \ldots, \frac{h^q y_n^{(q)}}{q!}\right]
\end{equation}
Here, $y_n$ is a vector of abundances at the current time, $h$ is the
current step-size (i.e., $h = t_{n+1} - t_n$), and $\dot{y}_n,
\ddot{y}_n, \ldots, y^{(q)}$ are the time derivatives of
$y_n$. To evolve the system from time $t_n$ to $t_{n+1}$, the
calculation can be divided into two steps: (i) the predictor step, and
(ii) the corrector step.

For the predictor step, the Nordsieck vector is used to predict the
future state of the system:
\begin{equation}
  z_{n+1}^{(0)} = z_n A(q)
  \label{eq:predictor}
\end{equation}
where $A(q)$ is a $(q+1) \times (q+1)$ matrix defined by
\begin{equation}
  \label{eq:correctorA}
  A^{ij}(q) = \left\{
  \begin{array}{l l}
    \begin{array}{c}\phantom{\left(\right.}0
    \end{array}&\quad \text{if } i < j \\
    \left(\begin{array}{c}i\\j
      \end{array}\right) = i!/j!(i-j)!&\quad \text{if } i \ge j
  \end{array} \right.
\end{equation}
Following this, the corrector step is taken. By defining a correction
vector, $e_n$, as that required to adjust the predicted
$y_{n+1}^{(0)}$ abundances to the final solution $y_{n+1}$, the
corrected Nordsieck vector can be corrected using:
\begin{equation}
  \label{eq:correctednord}
  z_{n+1} = z_{n+1}^{(0)} + e_{n+1} \ell.
\end{equation}
$\ell$ is computed using
\begin{equation}
  \sum_{j=0}^{q} \ell_j x^j = \prod_{i=1}^{q} (1+x/\xi_{i}), \label{eq:ell}
\end{equation}
where
\begin{eqnarray}
  x &=& (t-t_{n+1})/h \\
  \xi_i &=& (t_{n+1} - t_{n+1-i})/h.
\end{eqnarray}
In practice, these values are computed through a straightforward
iterative procedure~\citep{BYR75}.

The correction vector is calculated using Newton-Raphson
iteration. The procedure is to find the iterative correction,
$\Delta^{m}$, necessary to adjust each corrector step $m$ until
convergence reaches some predefined value. Therefore, we must solve:
\begin{eqnarray}
  \label{eq:newton}
  \left[ \mathbf{I} - \frac{h}{l_1}
        \mathbf{J}\right]\Delta^{(m)} &=& -(y_{n+1}^{(m)} -
      y_{n+1}^{(0)}) - \frac{h}{l_1} (f(y_{n+1}^{(m)}) -
      \dot{y}_{n+1}^{(0)}) \\
      y_{n+1}^{(m+1)} &=& y_{n+1}^{(m)} + \Delta^{(m)} 
\end{eqnarray}
Here, $f(y_{n+1}^{(m)})$ is the first time derivative of
$y_{n+1}^{(m)}$, and must be re-calculated at each corrector
iteration.  $\mathbf{J}$ is the Jacobian matrix, $\mathbf{J}
= \partial y/\partial t$, evaluated using the predicted values
calculated in Eq.~(\ref{eq:predictor}). In principle, this matrix
should be updated for each corrector iteration step, but provided that
the prediction is suitably close to the true solution, updating
$\mathbf{J}$ only once per time-step is necessary. Typically, after
only a few Newton-Raphson iterations, solution convergence is achieved
and Eq.~(\ref{eq:correctednord}) is used to correct the Nordsieck
vector.

While the method discussed so far can be used to solve numerical
problems, automatic step-size control cannot be achieved without some
estimation of the step error. Following successful convergence, the
error can be estimated using the Taylor series truncation error:
\begin{equation}
  \label{eq:error}
  E_{n+1}(q) = \frac{1}{l_1}\left[ 1+\prod_{i=2}^{q}\left(
      \frac{t_{n+1} - t_{n+1-i}}{t_{n-1}-t_{n+1-i}} \right) 
  \right]^{-1} e_{n+1}.
\end{equation}
Equation~(\ref{eq:error}) can now be used to estimate the step-size to
be used in the following step. If $\epsilon$ is defined as the error
tolerance per step, then the new step-size is calculated in an
analogous way to Eq.~(\ref{eq:bader-newstepsize}) by
\begin{equation}
  \label{eq:stepsize}
  h' = h \left( \frac{\epsilon}{\textrm{max}(E_{n+1}(q))} \right)^{1/q+1}.
\end{equation}
This is usually limited in practice by multiplying a conservative
factor of 0.25 and by preventing large changes in the
step-size~\citep{BYR75}. In the same way as discussed for the
Bader-Deuflehard method, relative errors are calculated by scaling the
errors by their corresponding abundances for those larger than a
pre-defined value $y_{\text{scale}}$, as in Eq.~(\ref{eq:yscale}).

Beyond step-size estimation, the order of the method, $q$, can also be
altered to automatically select the most efficient method. Here, we
follow the recommendation of \cite{BYR75} and allow only order changes
of $q \pm 1$. Every so often (i.e., after at least $q$ steps at the
current order), trial error estimates are calculated for increasing
and decreasing order:
\begin{eqnarray}
  E_{n+1}(q-1) &=& -\left[ \frac{\xi_1 \xi_2
      \ldots \xi_{q-1}}{\ell_1(q-1)}\right] \frac{h^q
    y_{n+1}^{(q)}}{q!} \label{eq:errorqchangep}\\
  E_{n+1}(q+1) &=& \frac{-\xi_{q+1} (e_{n+1} - Q_{n+1}e_{n})}{
    (q+2)\ell_1(q+1)\left[1+\prod_2^q \frac{t_{n+1}-t_{n+1-i}}{t_n -
      t_{n+1-i}} \right]} \label{eq:errorqchangem}
\end{eqnarray}
where
\begin{eqnarray}
  Q_{n+1} &=&
  \frac{C_{n+1}}{C_n}\left(\frac{h_{n+1}}{h_n}\right)^{q+1} \label{eq:Qeen}\\
  C_{n+1} &=& \frac{\xi_1 \xi_2 \ldots \xi_{q}}{(q+1)!} 
  \left[ 1+\prod_2^q \frac{t_{n+1}-t_{n+1-i}}{t_n - t_{n+1-i}} \right] 
  \label{eq:Ceen}
\end{eqnarray}
By using these expressions, trial step-sizes are found from
Eq.~(\ref{eq:stepsize}). The largest step-size is then chosen and the
order altered accordingly if necessary. One final calculation is
necessary before the next time-step can be taken, and that is to scale
the Nordsieck vector with the new step-size and order. Order scaling
is first applied. If the order is to increase or remain constant, the
Nordsieck vector requires no scaling. However, for reducing order, a
factor $\Delta_i$ must be subtracted from each column in $z_{n+1,i}$,
where
\begin{eqnarray}
  \label{eq:reduceord}
  \Delta_i &=& d_i z_{n+1,q} \\
  d_i &=& x_i^2 \prod_j^{q-2}(x_i - \xi_j) 
\end{eqnarray}

Regardless of the order change, the Nordsieck vector must finally be
scaled to the new step-size with $\eta = h'/h$:
\begin{equation}
  \label{eq:zscale}
  z_{n+1}' = z_{n+1} \textrm{diag}(1,\eta,\eta^2,\ldots,\eta^q)
\end{equation}
Finally, once all of these steps have been completed, the code returns
to Eq.~(\ref{eq:predictor}) and the next step is computed. This method
automatically takes care of starting steps, by using $q=1$ until
enough history has been built up to utilise higher orders. It is fully
automatic and although complicated, reduces the need for expensive
Newton-Raphson iteration by making beneficial use of past behaviour of
the system.

\section{Input physics and test cases}
\label{sec:tests}

To investigate the efficiency of each of the methods described in
Sec.~\ref{sec:methods}, a test suite of reaction networks and
post-processing profiles is used. The purpose of these different cases
is to investigate the performance of the integration schemes in a
variety of nucleosynthesis post-processing situations. Different
environments exhibiting different properties such as rapid changes of
temperature or number of reactions involved are useful in highlighting
the advantages and disadvantages of the methods described above.

\subsection{Temperature and density profiles}
\label{sec:profiles}

Two flat profiles are used to explore the behaviour of the integration
methods discussed here when no step-size limits related to profile
behaviour are necessary. The first, referred to here as the ``Flat''
profile, is the same profile used by \cite{TIM99}, i.e., $T=3 \times
10^9$~K and $\rho=2 \times 10^8$~\gcm\ for 1~s. The second profile
``Flat2'' is a less extreme profile that is more applicable to novae,
AGB stars, etc. In this case, the temperature and density are $T=3
\times 10^8$~K and $\rho=2 \times 10^5$~\gcm\ for 100~s.

\begin{figure}
  \centering
  \includegraphics[width=0.8\hsize]{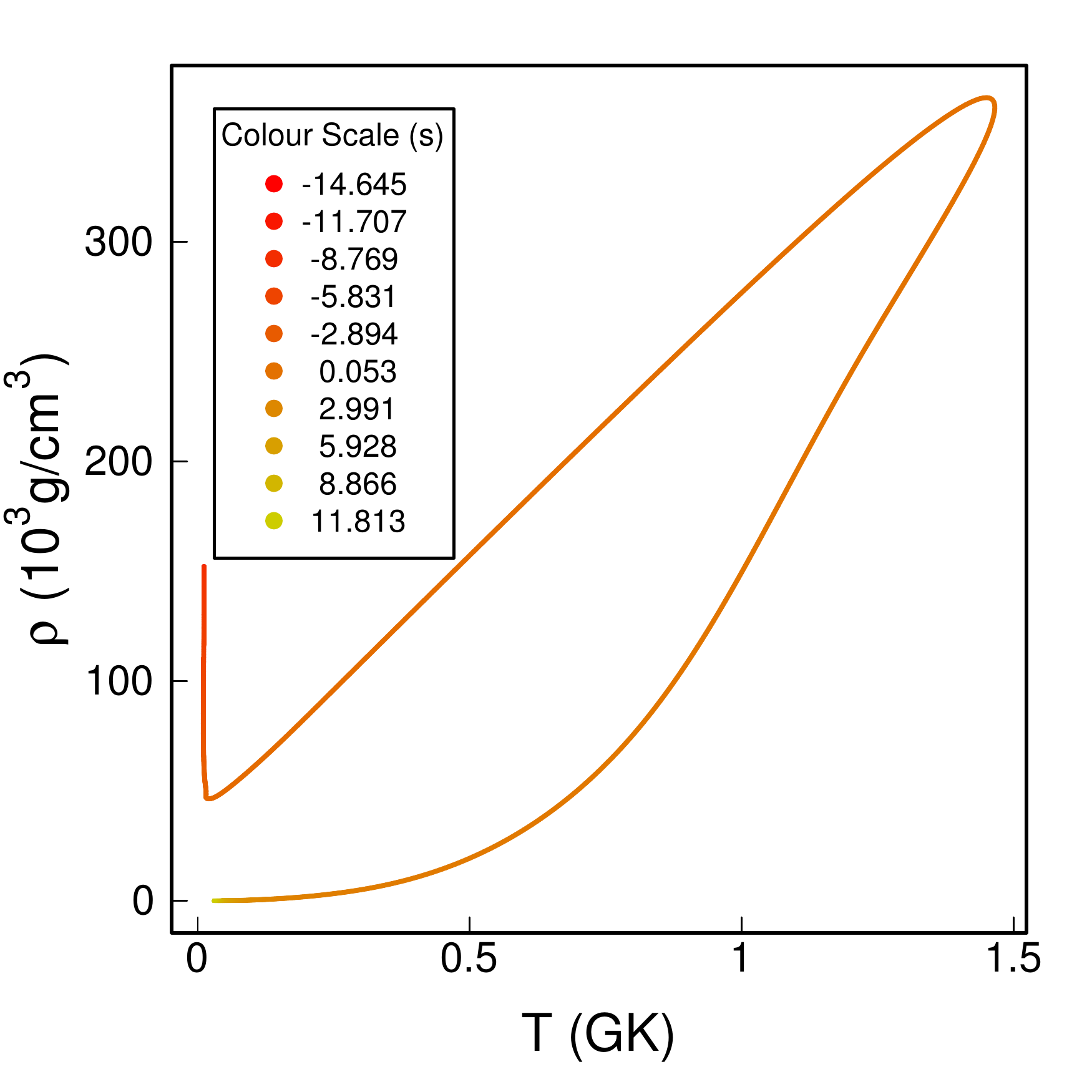}
  \caption{(Colour online) Merger profile. Shown is the density
    plotted against the temperature with time represented by colour,
    with red for early times; yellow for late times. Time is labelled
    with respect to peak temperature, at which time, $t=0$~s. Most of
    the nucleosynthesis is expected to occur, in this particular case,
    close to the start of the profile and within a 30~s window.}
  \label{fig:merger}
\end{figure}

The merger profile is extracted from Smoothed Particle Hydrodynamic
(SPH) simulations\ \citep[see, for example,][]{LOR10} of white dwarf
mergers and is shown in Fig.~\ref{fig:merger}. For this particular
model, the profile corresponds to a particular tracer particle in the
merging of a $0.4 M_{\odot}$ helium white dwarf with a $0.8 M_{\odot}$
carbon-oxygen white dwarf. For this profile, the tracer particle was
identified to exhibit significant nuclear reaction activity, and is
therefore ideally suited for the present study. More details of this
particular case can be found in \cite{LON11}. The profile follows a
characteristic shape. Initially, as material leaves the Roche lobe of
the secondary white dwarf, the density and temperature drop. The
material reaches and impacts the primary white dwarf's surface (after
about 14~s), undergoes rapid heating to $T \approx 1.5 \times 10^9$~K
in just a few seconds, and undergoes subsequent cooling on a timescale
of around 10 seconds. The remainder of the profile is relatively cool,
in which no nuclear processing occurs. This profile represents a
post-processing situation in which the allowed time-step in
nucleosynthesis integration is limited by the profile shape at early
times, and not necessarily by theoretically allowed step-sizes
computed by the integration method.

\begin{figure}
  \centering
  \includegraphics[width=0.8\hsize]{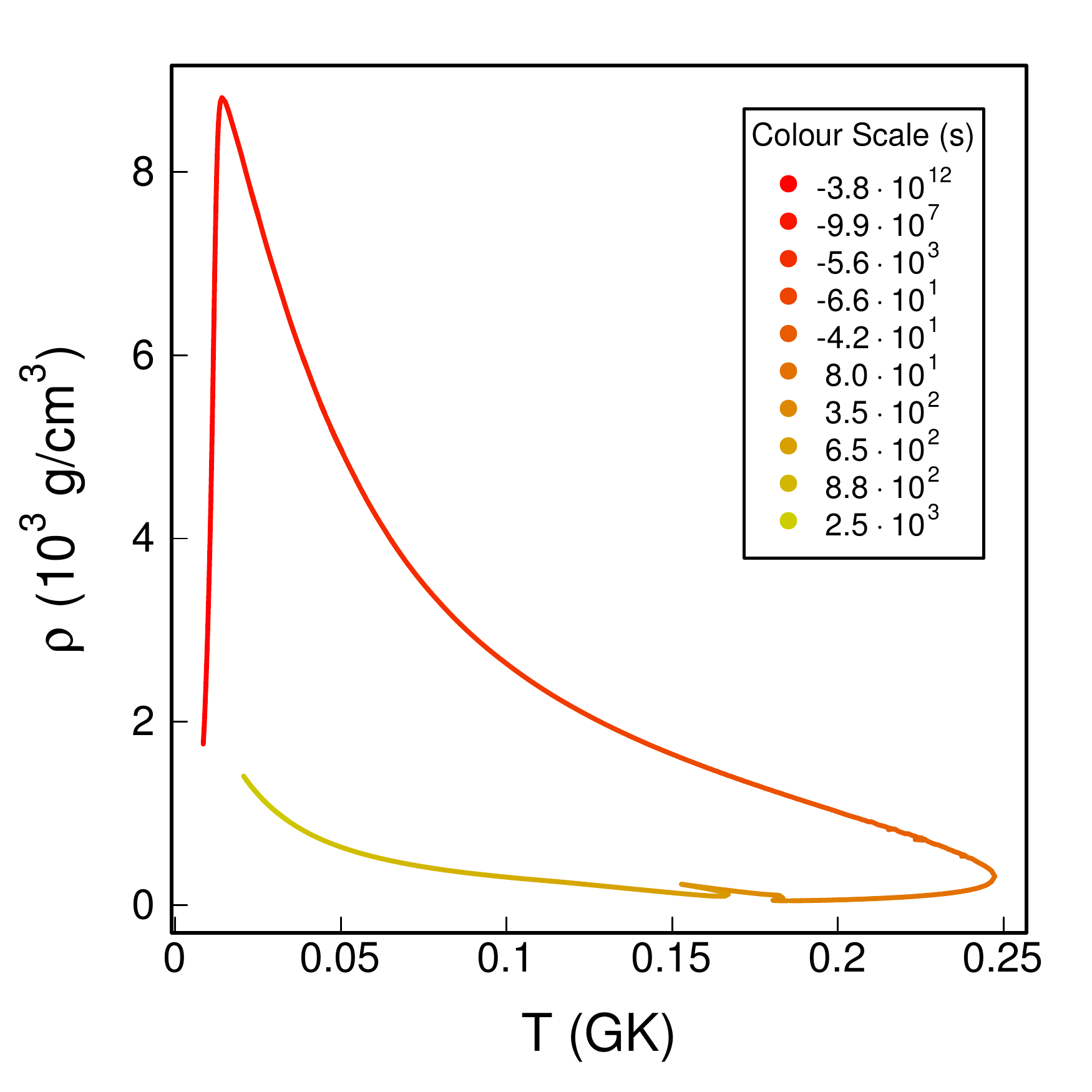}
  \caption{(Colour online) Same as Fig. \ref{fig:merger}, but for the
    nova profile. This highlights the fact that the profile is
    relatively flat for most of the evolutionary history, with most of
    the nucleosynthesis occurring in a few thousand second window.}
  \label{fig:nova}
\end{figure}

To investigate the behaviour of the integration methods in nova
nucleosynthesis, we use the temperature and density versus time
profile of the innermost envelope shell, computed with the multi-zone
1-D hydrodynamic code ``SHIVA''~\citep{JOS99}. This model comprises of
a $1.25$~M$_{\odot}$ ONe white dwarf accreting matter with solar
metalicity at a rate of $\dot{M}_{\text{acc}} = 2 \times
10^{-10}$~M$_{\odot}\,\text{yr}^{-1}$. The profile, displayed in
Fig.~\ref{fig:nova}, was also used in the sensitivity study performed
by \cite{ILI02}. As material is accreted from the companion star onto
the white dwarf, density gradually increases on a long timescale
($\sim 10^{12}$~s) until conditions are reached in which explosive
ignition of the material can occur. At that point, nuclear burning
under degenerate conditions causes a rapid increase in temperature
until degeneracy is lifted and the material subsequently expands and
cools to settle at the end of the profile. This outburst stage occurs
on a very short timescale ($\sim 1000$~s) compared to the quiescent
accretion phase of the profile. It is therefore essential to utilise
an integration method that is capable of adaptive step-size control
over a very large range of time-scales. Furthermore, the method must
be capable of detecting loss of accuracy coming from sudden changes in
nucleosynthesis, as we discuss in more detail later.

\begin{figure}
  \centering
  \includegraphics[width=0.8\hsize]{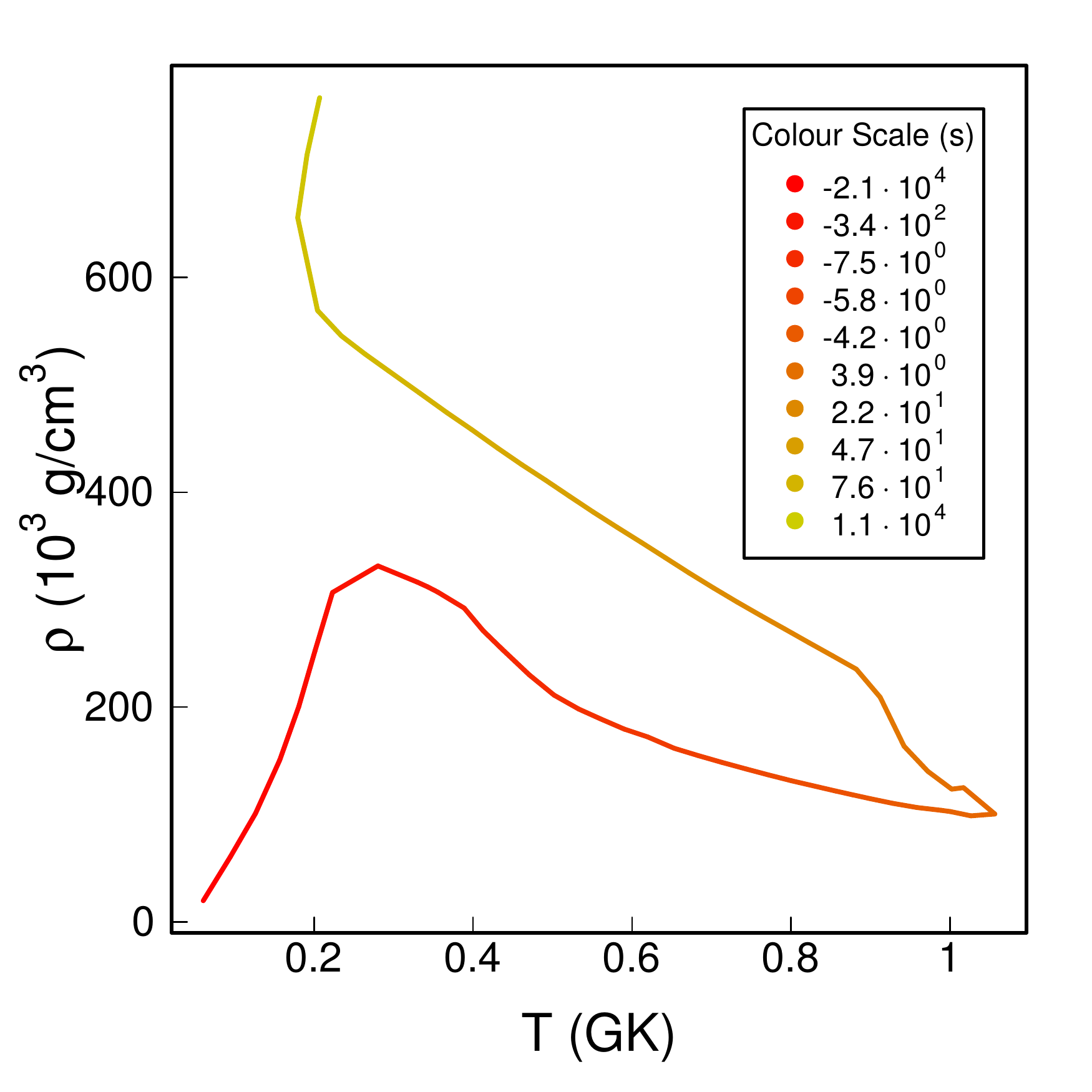}
  \caption{(Colour online) Same as Fig.~\ref{fig:merger} but for the
    X-ray burst profile.}
  \label{fig:xrb}
\end{figure}

Similarly, for X-ray burst nucleosynthesis, a temperature and density
versus time profile is used from the innermost shell computed by the
1-D SHIVA hydrodynamic code \citep{JOS10} and shown in
Fig.~\ref{fig:xrb}. The burst is driven by accretion of solar
metalicity material onto the surface of a $1.4$~M$_{\odot}$ neutron
star at a rate of $\dot{M}_{\text{acc}} = 1.75 \times
10^{-9}$~M$_{\odot}\,\text{yr}^{-1}$.  Similarly to the nova model
discussed previously, for most of the profile, material undergoes
compression and heating until conditions are reached at which nuclear
burning begins after $\sim 10^4$~s. At this time, rapid heating and
subsequent cooling of the material occurs in a 10~s window, followed
by a quiescent settling period for a further $\sim 10^4$~s. This
dramatic range of time-scales and conditions serves as a challenge to
the network integration methods considered here.

\begin{figure}
  \centering
  \includegraphics[width=0.8\hsize]{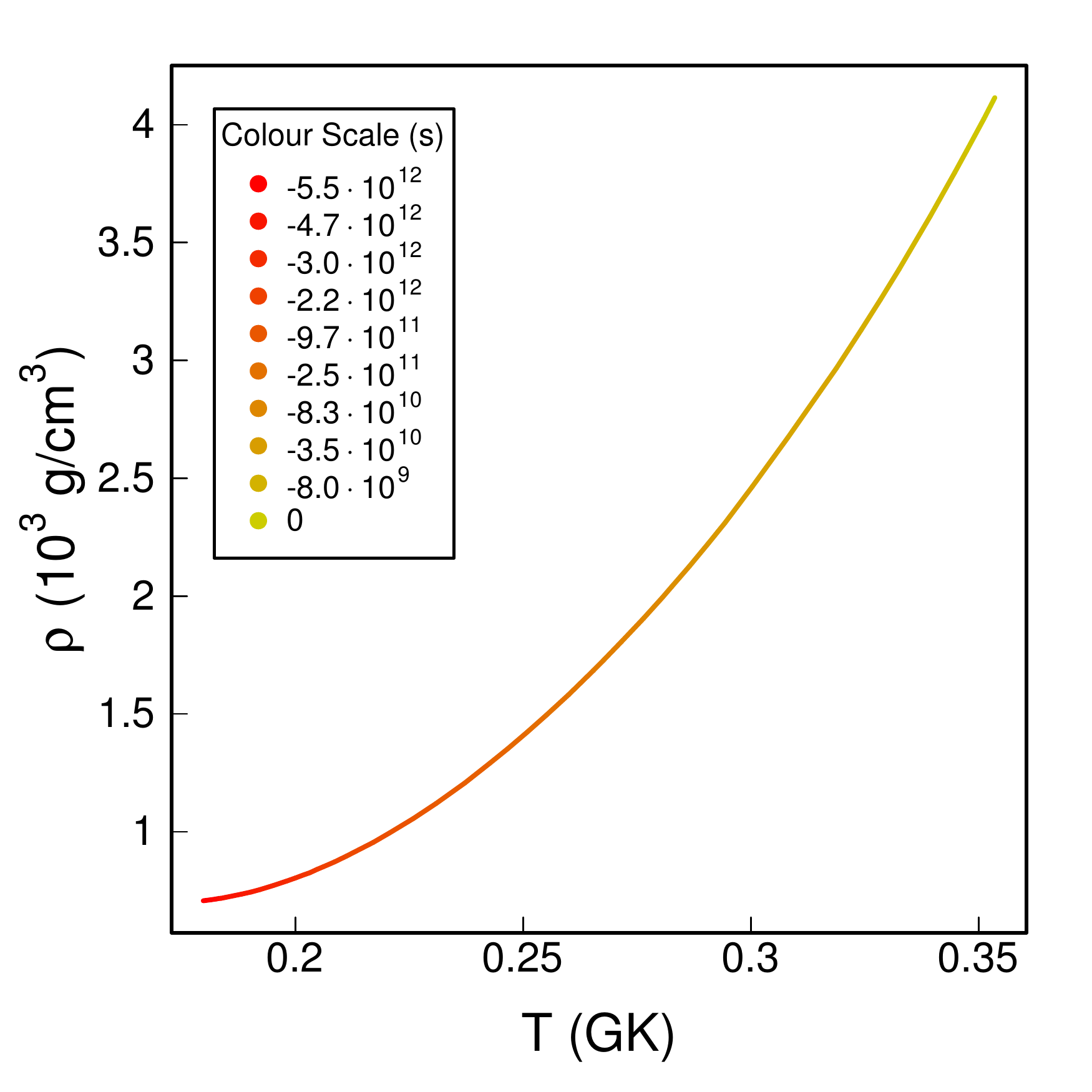}
  \caption{(Colour online) Same as Fig.~\ref{fig:merger} but for the
    s-Process profile. The times are scaled to the time at which
    helium is exhausted in the core. Once again, we see evolution
    approach the more complex nucleosynthesis region only for a
    relatively brief period at the end of the profile.}
  \label{fig:sprocess}
\end{figure}

Finally, in order to characterise the integration methods for a range
of scenarios, an s-process profile (shown in Fig.~\ref{fig:sprocess})
is used. The profile is extracted from the 1-D hydrostatic core helium
burning models of \cite{THE00}, which follow the evolution of a
25~$M_{\odot}$ star until helium is exhausted in the core. The profile
is smoother than those of the nova and X-ray burst profiles and thus
will probe a different numerical nucleosynthesis regime. The
conditions over much of the profile are sufficient for helium burning
to occur, but the s-process only becomes active at the end of the
profile when the \Nepa\ neutron source is activated. This final
profile provides a good test of situations for which there is
continuous processing of material that changes in nature over time as
opposed to the brief bursts of activity that are characterised by the
nova, merger, and X-ray burst models.

\subsection{The networks}
\label{sec:networks}

Reaction rates in the present work are adopted from the
\texttt{starlib} reaction rate library~\citep[summarised in][]{SAL13}
for each network considered here. Each reaction rate is tabulated on a
grid of temperatures between $T=10$~MK and $T=10$~GK. Cubic spline
interpolation is used to compute the rate between temperature grid
points. Since small nuclear networks do not present much of a
challenge for modern computers, we only consider more detailed
reaction networks here.

A large, 980 nucleus network (dubbed the ``980'' network in the
following analysis) suitable for massive star nucleosynthesis is
considered for both flat profiles discussed in
Sec~\ref{sec:profiles}. This network is based on the network presented
by \cite{WOO95}, but extended to barium. The network contains 9841
reactions linking nuclei from hydrogen to barium ($Z=56$) with a 100\%
\nuc{4}{He} initial composition. This represents a computationally
more intensive network, requiring LU-decomposition of a $980 \times
980$ Jacobian matrix at every time-step.

The profile used to follow nucleosynthesis in merging white dwarfs is
labelled the ``Merger'' network. This network is designed to allow for
detailed nucleosynthesis studies from hydrogen to germanium ($Z=32$),
with a total of 328 isotopes and 3494 reactions. The initial mass
fractions used in this particular network of \nuc{1}{H},
\nuc{4}{He}, and \nuc{3}{He} are 0.5, 0.5, and $1\times 10^{-5}$,
respectively. These initial abundances correspond to helium white
dwarf buffer regions that are necessary to explain abundances in some
hydrogen-poor stars~\citep[see][and references
therein]{LON11,JEF11}. 

The ``Nova'' network is based on that used by \cite{ILI02} with
initial abundances adopted from the JCH2 model in their Tab. 2. The
network comprises of nuclei from hydrogen to calcium (145 isotopes)
and 1274 reactions that allow for all common processes and their
reverse reactions. Similarly, the ``X-ray Burst'' network is adopted
from \cite{PAR08}. It contains 813 nuclei up to xenon, linked by 8484
reactions. Initial abundances are defined for 89 of these nuclei
similarly to those in \cite{PAR09}. Both of these networks are
modified only in that they include updated reaction rates from the
\texttt{starlib} database.

Finally, the ``s-Process'' network is designed for studying the weak
s-process in massive stars. A total of 341 nuclei are used from
hydrogen to molybdenum, linked by 3394 reactions. The initial
abundances are adopted from \cite{THE00}, and correspond to the
expected abundances prior to helium burning in a $25 M_{\odot}$ star.

\section{Results}
\label{sec:result}

\subsection{Integration times}
\label{sec:varying-parameters}

To fully characterise the integration methods for the profiles and
networks outlined in Sec.~\ref{sec:tests}, the variables impacting
integration accuracy constraints discussed in Sec.~\ref{sec:methods}
are varied. Gear's Method and the Bader-Deuflehard method both use
similar variables for controlling integration accuracy:
$y_{\text{scale}}$ and $\epsilon$. The first of these controls the
normalisation of abundances when calculating integration errors
through Eqs.~(\ref{eq:bader-newstepsize}) and~(\ref{eq:stepsize}).
Relative errors are used to improve the accuracy of the integration
routines for lower abundances. These are obtained by scaling
truncation errors by the abundance for each isotope. The minimum
abundance that this procedure is applied to is the integration
variable $y_{\text{scale}}$ through Eq.~(\ref{eq:yscale}). Scaling the
abundances in this way results in a gradual decline in weight as
abundances drop below this value, resulting in more stable integration
performance.

No analog of these variables exists for Wagoner's method in its
present form. Historically, the accuracy of any one integration result
is identified in an ad-hoc method that involves changing integration
variables until the abundances of interest converge to stable
values. Nevertheless, there are two variables, $y_{\text{t,min}}$ and
$S_{\text{scale}}$, that are generally good indicators of the
integration accuracy. $y_{\text{t,min}}$ controls the abundance below
which no error checking is performed. Note that this is different from
the $y_{\text{scale}}$ above, which still allows the method to use low
abundance isotopes, but with less weight. In this case,
$y_{\text{t,min}}$ represents a sharp cut-off. $S_{\text{scale}}$ is a
simple parameter that limits the maximum allowable step-size (i.e.,
larger values correspond to smaller time steps). This variable cannot
be easily compared across usage situations, but within a single
application, it can be used for characterising the method and
guaranteeing that no sudden profile changes are missed by the
integration method.

All tests were computed on a desktop PC with an 8-core Intel
  Core i7 2.67~GHz CPU. Only one core was utilised for all tests,
  ensuring that additional processes were kept to a minimum. However,
  all computation times presented here are normalised to the Wagoner's
  method calculations with \SSC$=1000$ and \YTMIN$=10^{-12}$ to remove
  most architecture specific computation dependence.
Figures~\ref{fig:times-flat} to \ref{fig:times-sprocess} show the
total integration time required to compute nucleosynthesis for each of
the test cases presented here as integration parameters are varied. In
our integration method runs, we impose a $100\,000$ step
limit on each test. This is to ensure that the tests are completed in
a reasonable amount of time. When a method exceeds this limit, the
approximate integration time can be inferred by extrapolating the
curves.

\subsection{Results convergence}
\label{sec:result-convergence}

\begin{figure}
  \centering
  \includegraphics[width=0.7\hsize]{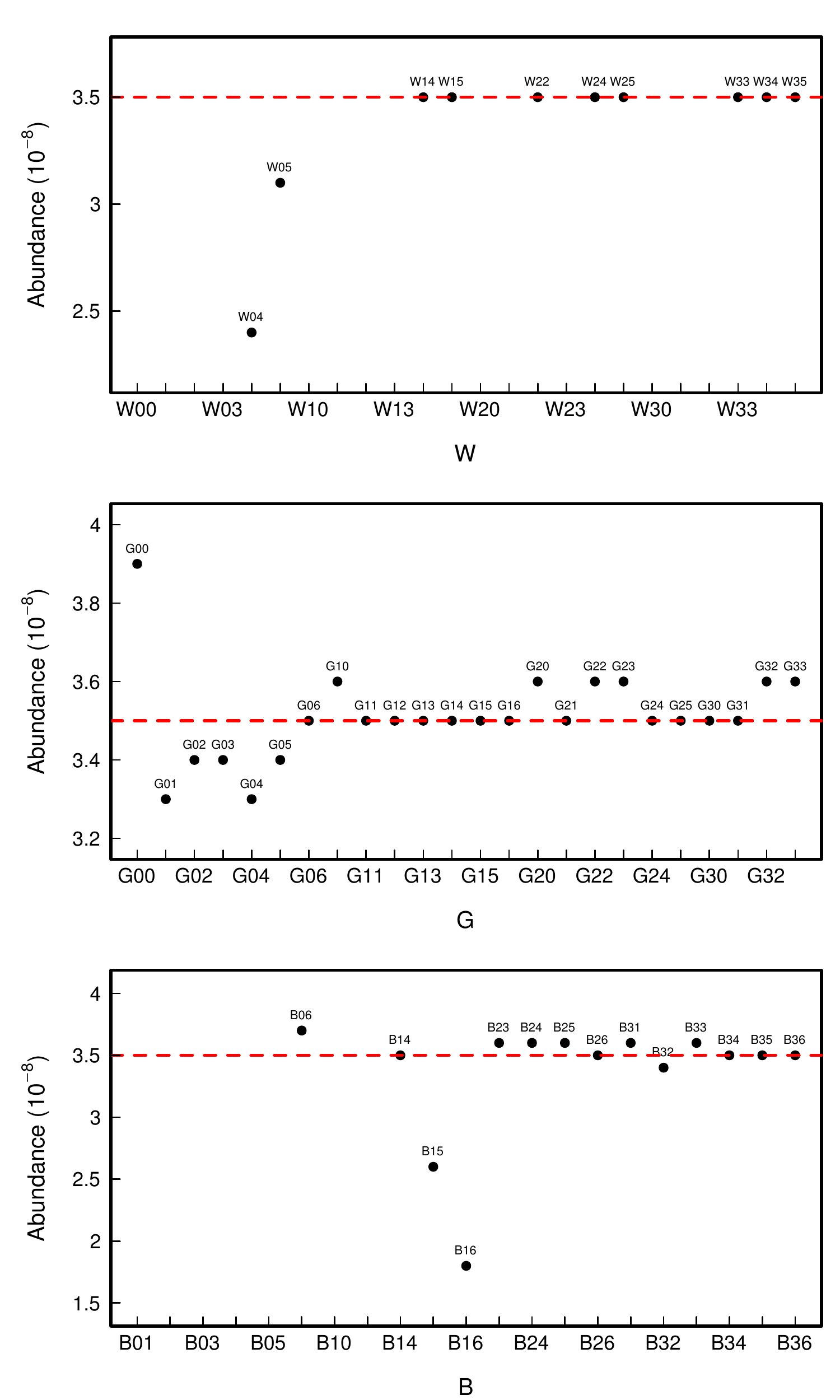}
  \caption{Example of convergence tests for the 'Nova' test case. See
    Tab.~\ref{tab:pars} for label interpretation. For example,
      point `G12' in the middle panel corresponds to a Gear's method
      run with integration parameters of \EPS$=10^{-3}$ and
      \YSCALE$=10^{-8}$.}
  \label{fig:Converge}
\end{figure}

Comparing integration accuracy input parameters between methods is
challenging. Using the same parameters for two different methods can,
particularly for less constrained cases, lead to different
accuracy in final results\footnote{For the purpose of the tests
  performed here, we evaluate accuracy of results by comparing them
  with the most constrained calculations. Those have been carefully
  evaluated to ensure that convergence to stable results has been
  achieved for every isotope included in the network.}. To account for
this, we have performed additional tests as follows. Initially, a
species in the network is chosen at the desired final abundance. For
the following example, we consider the final abundance of \nuc{19}{F}
in the Nova test case, whose final abundance should be $3.5\times
10^{-8}$. The convergence behaviour of each method was then
investigated according to the integration parameters outlined in
Tab.~\ref{tab:pars}. Results of the convergence tests are shown in
Fig.~\ref{fig:Converge} for the 'Nova' profile and network.

It is clear from Fig.~\ref{fig:Converge} that accurate results at the
$X \sim 10^{-8}$ level are not obtained for all integration parameters
(many runs for the Wagoner method achieve results far outside the
plotted region). For example, the `W04' run, corresponding to a
Wagoner's method run with \SSC$=10$ and \YTMIN$=10^{-12}$ (see
Tab.~\ref{tab:pars} for label interpretation) did not obtain an
accurate result, with an error of about 30\%. This procedure was
repeated with varying precision requirements from $X = 10^{-4}$ to $X
= 10^{-12}$ for all test cases considered here. For example, although
test B14 exhibits converged results for this test case, it was not
consistently accurate in others. By combining the results for all test
cases, we obtained robust choices of representative integration
parameters, which we consider to be `safe' for the purposes of this
work. The representative integration parameters found are (i)
\SSC$=1000$, \YTMIN$=10^{-12}$ for Wagoner's method; (ii)
\EPS$=10^{-3}$, \YSCALE$=10^{-10}$ for Gear's method; and (iii)
\EPS$=10^{-5}$, \YSCALE$=10^{-15}$ for the Bader-Deuflehard method.

\begin{table}
  \caption{Integration parameter reference table.}
  \centering
  \begin{tabular}{c||cc|cc|cc}
    \hline \hline
    & \multicolumn{2}{c|}{Wagoner} & \multicolumn{2}{c|}{Gear} & \multicolumn{2}{c}{Bader-Deuflehard} \\ 
    Index & \SSC & \YTMIN & \EPS & \YSCALE & \EPS & \YSCALE \\ \hline
    0 & 10    & $ 10^{-3}$  & $10^{-1}$ & $10^{-3}$ &  $10^{-1}$ & $10^{-3}$  \\ 
    1 & 100   & $ 10^{-6}$  & $10^{-3}$ & $10^{-6}$ &  $10^{-3}$ & $10^{-6}$  \\ 
    2 & 1000  & $ 10^{-8}$  & $10^{-5}$ & $10^{-8}$ &  $10^{-5}$ & $10^{-8}$  \\ 
    3 & 10000 & $ 10^{-10}$ & $10^{-6}$ & $10^{-10}$ & $10^{-6}$ & $10^{-10}$ \\ 
    4 &       & $ 10^{-12}$ &           & $10^{-12}$ &           & $10^{-12}$ \\ 
    5 &       & $ 10^{-15}$ &           & $10^{-15}$ &           & $10^{-15}$ \\ 
    6 &       & $ 10^{-18}$ &           & $10^{-18}$ &           & $10^{-18}$ \\
    \hline \hline
  \end{tabular}
  \tablefoot{Usage example: point
    `B24' in Fig.~\ref{fig:Converge} corresponds to a Bader-Deuflehard method
    run with integration parameters of
    \EPS$=10^{-5}$ and \YSCALE$=10^{-12}$.}
  \label{tab:pars}
\end{table}

\subsection{Bottleneck identification}
\label{sec:bottleneck-identification}

It can also be illustrative to consider the fraction of time required
by the most computationally intensive components of the calculations:
(i) computation of the Jacobian matrix, which includes interpolating
reaction rate tables; (ii) overhead required by the integration
method; and (iii) linear algebra solution, which is dominated by
matrix LU-decomposition as discussed in Sec.~\ref{sec:methods}. The
fraction of computation time used by each of these components was
computed for representative integration variables for each
method. The time-fractions required by each of these computation components
are presented in Tab. \ref{tab:timefractions}. These were computed
using the set of standard integration parameters found to provide
comparable accuracy between methods. All absolute times have been
normalised to the Wagoner's method runs.

\begin{figure}
  \centering
  \includegraphics[width=0.8\hsize]{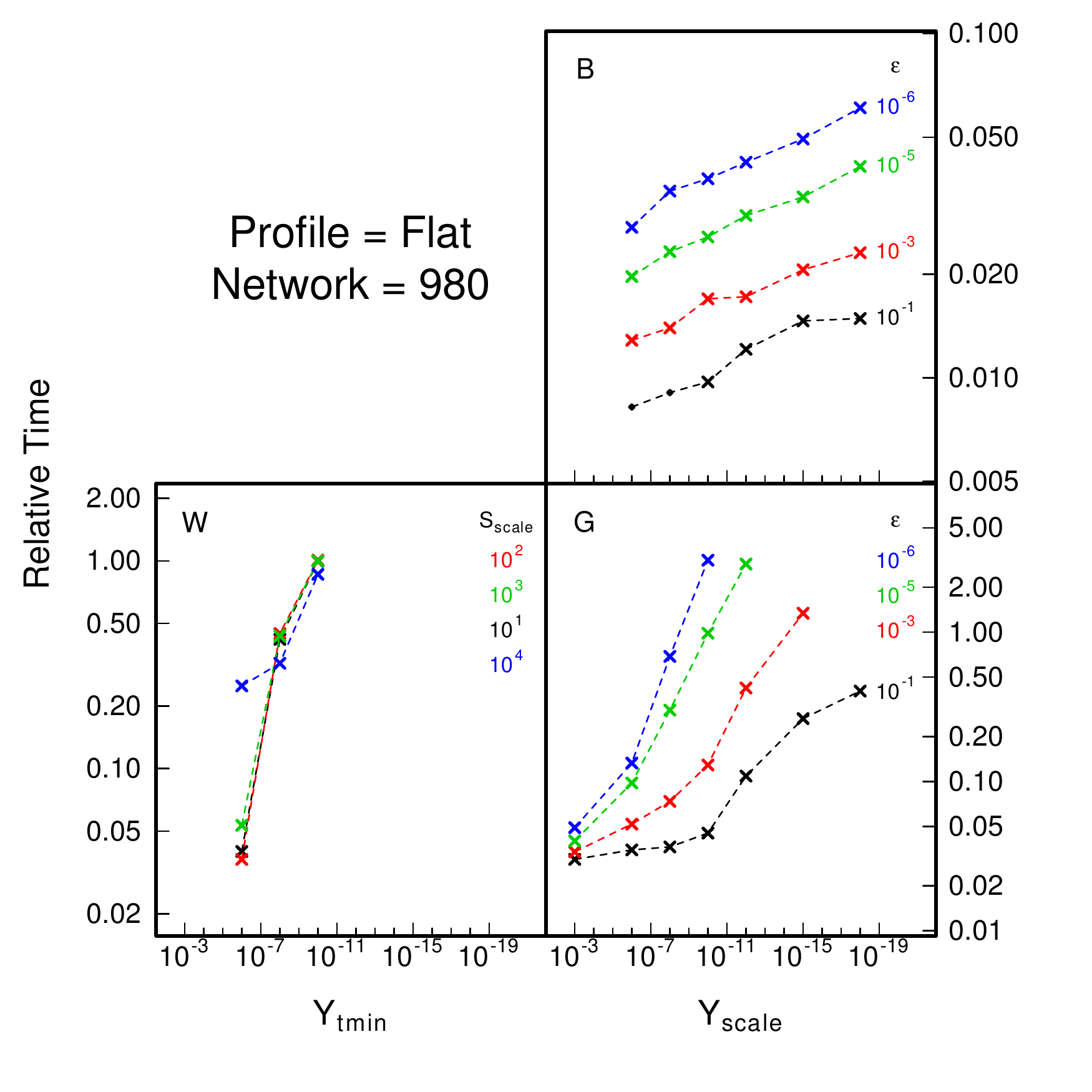}
  \caption{ (Colour online) Run-times computed for case 1 - ‘Flat’ and
    normalised to the Wagoner's method calculation with
    $S_{\text{scale}}=1000$ and $y_{\text{t,min}}=10^{-12}$ (see
    text). Points indicated with a cross represent calculations that
    completed successfully with abundances that meet the accuracy
    criteria outlined in the text. Missing data points represent runs
    that did not successfully run, either arising from convergence
    problems within the code or from exceeding the maximum number of
    time-steps allowed. The labels “W”, “B”, and “G” represent
    run-times for Wagoner’s, the Bader-Deuflehard, and Gear’s methods,
    respectively. }
  \label{fig:times-flat}
\end{figure}

\begin{figure}
  \centering
  \includegraphics[width=0.8\hsize]{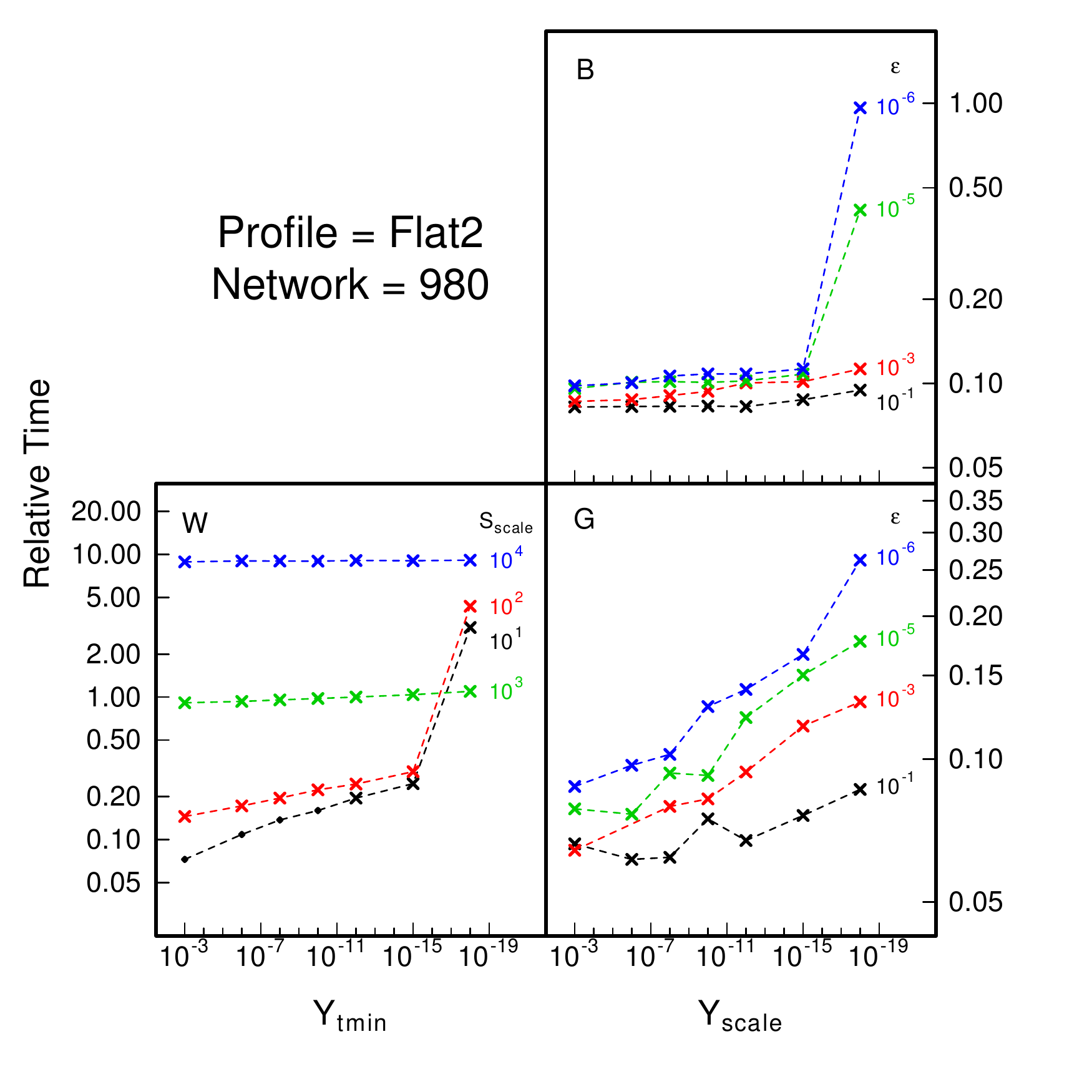}
  \caption{(Colour online) Same as Fig.~\ref{fig:times-flat} but for
    case 2 - the `Flat2' profile. Additionally, a closed circle symbol
    is used to represent runs that completed successfully but with
    results that do not meet the accuracy criteria outlined in the
    text. For example, the Wagoner's method run for
    $S_{\text{scale}}=10$ and $Y_{t,min}=10^{-1}$ reports successful
    completion, but computes an abundance for \nuc{4}{He} that exceeds
    the correct result by a factor of roughly 10.}
  \label{fig:times-flat2}
\end{figure}

\begin{figure}
  \centering
  \includegraphics[width=0.8\hsize]{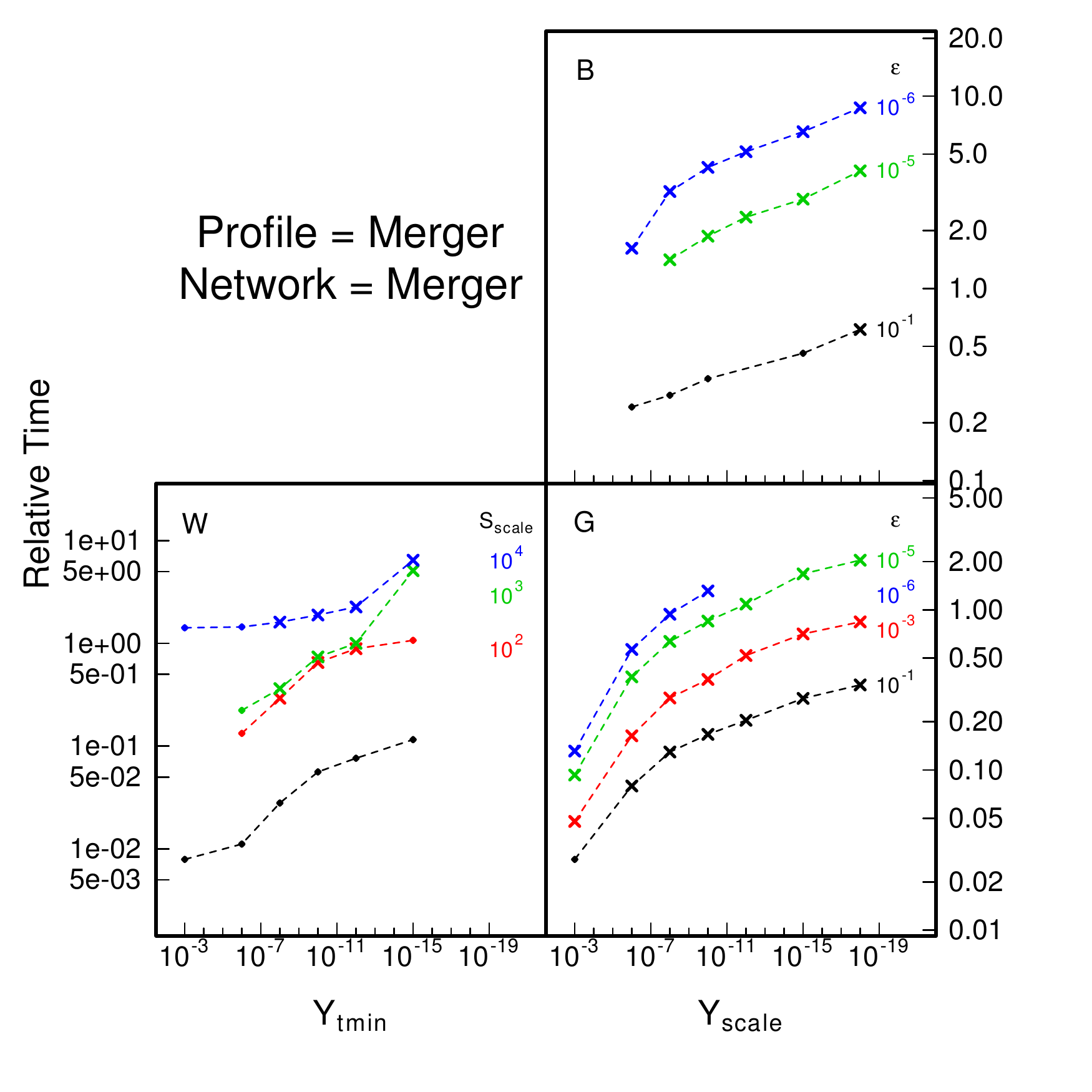}
  \caption{(Colour online) Same as Fig.~\ref{fig:times-flat} but for
    case 3 - the `Merger' profile.}
  \label{fig:times-merger}
\end{figure}

\begin{figure}
  \centering
  \includegraphics[width=0.8\hsize]{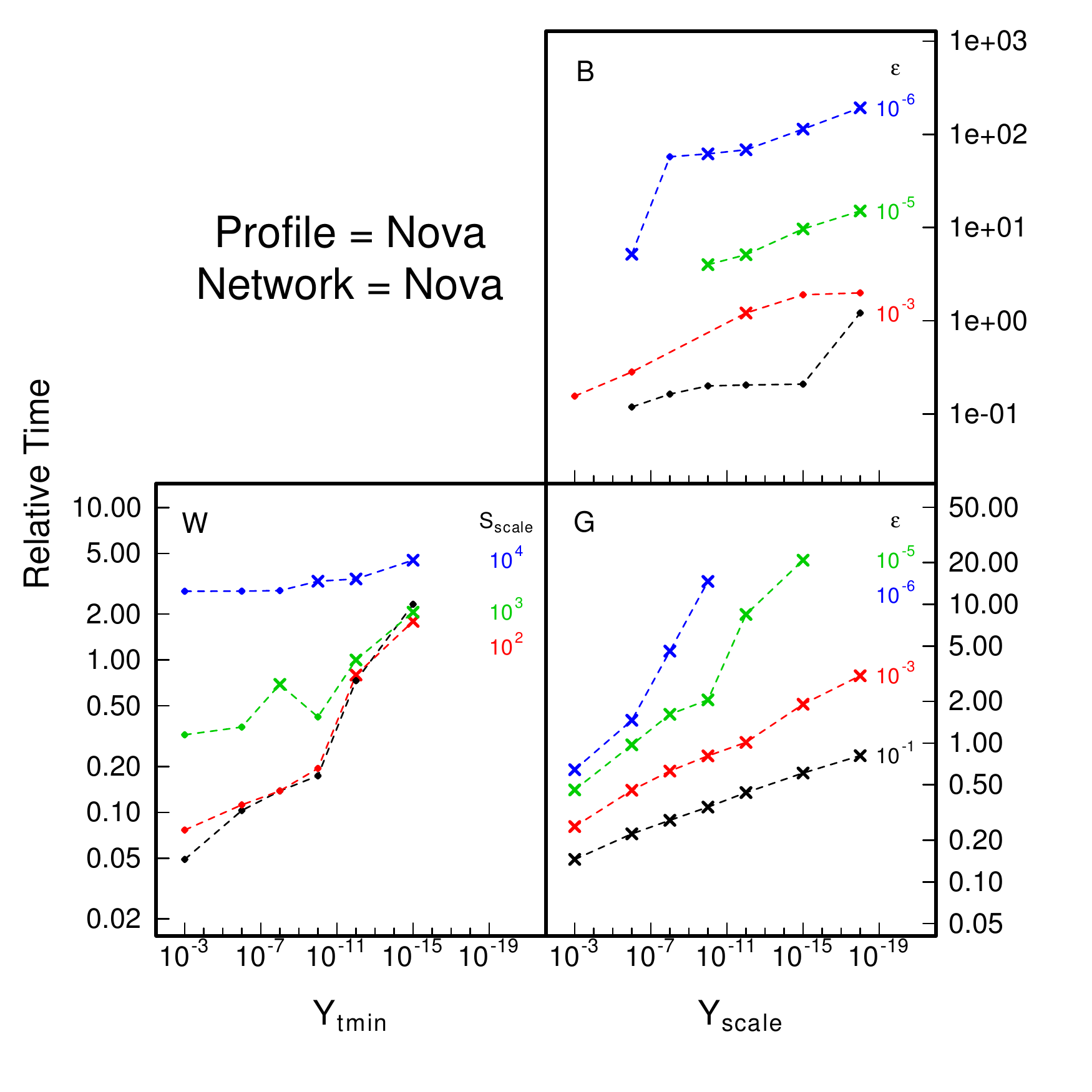}
  \caption{(Colour online) Same as Fig.~\ref{fig:times-flat} but for
    case 4 - the `Nova' profile.}
  \label{fig:times-nova}
\end{figure}

\begin{figure}
  \centering
  \includegraphics[width=0.8\hsize]{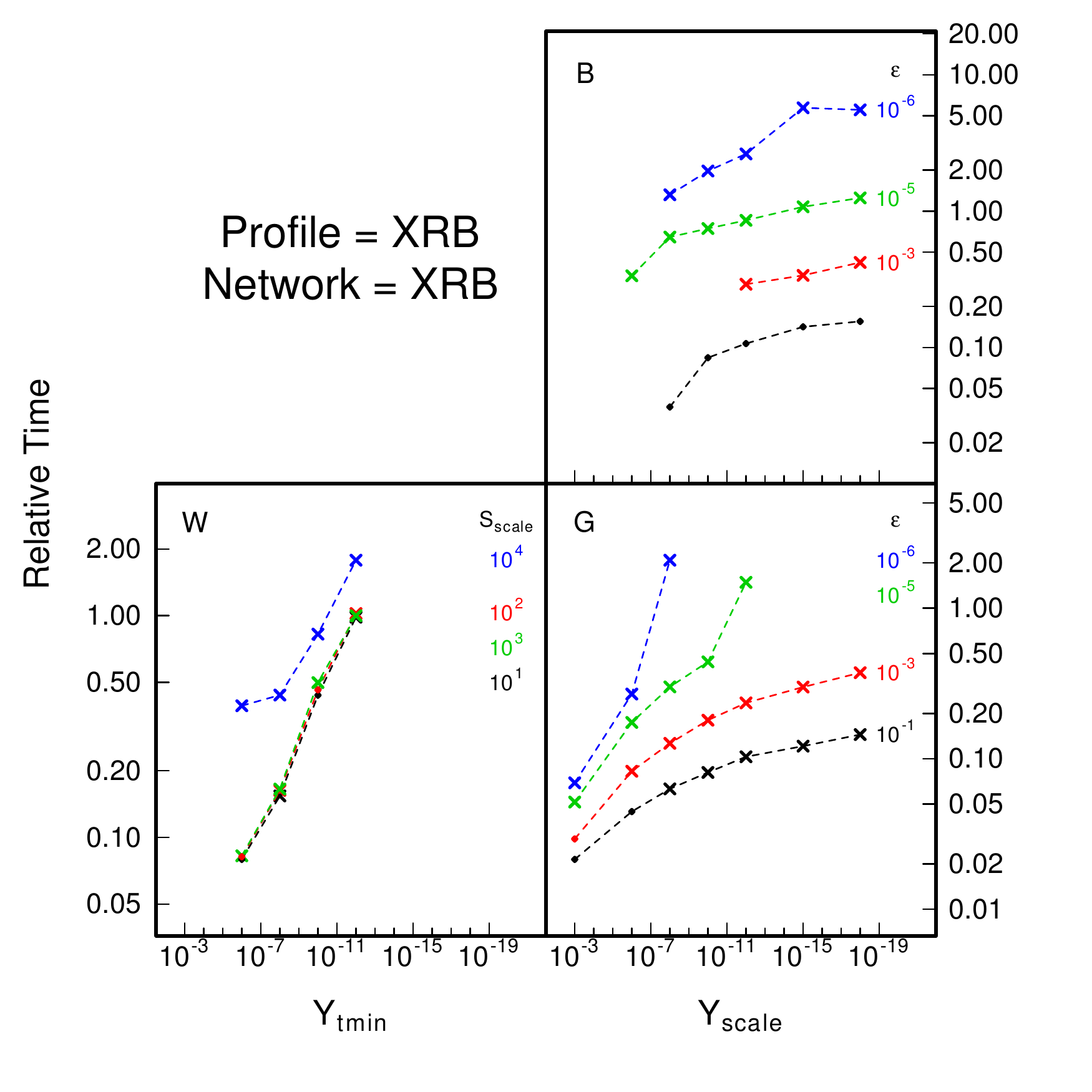}
  \caption{(Colour online) Same as Fig.~\ref{fig:times-flat} but for
    case 5 - the `X-ray Burst' profile.}
  \label{fig:times-xrb}
\end{figure}

\begin{figure}
  \centering
  \includegraphics[width=0.8\hsize]{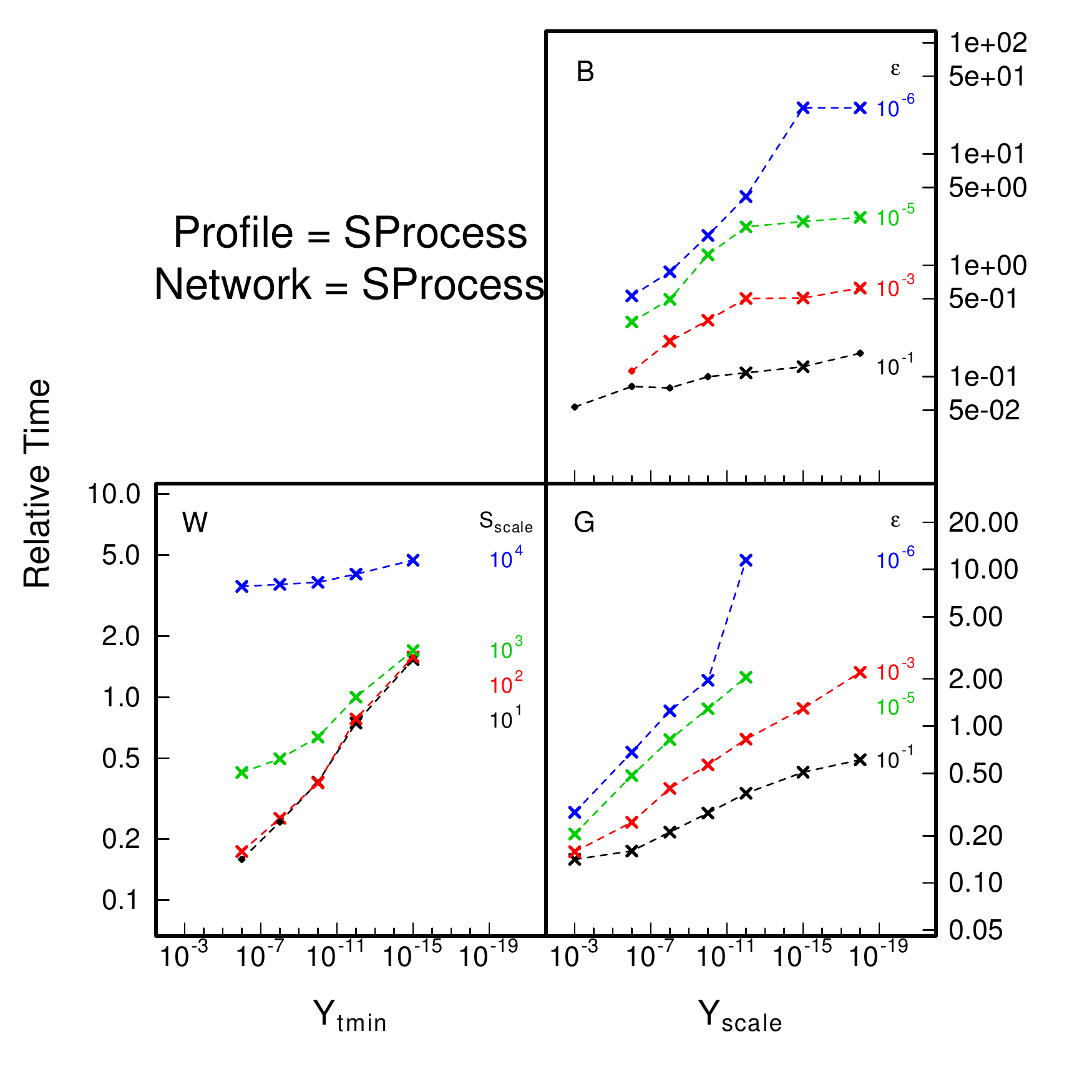}
  \caption{(Colour online) Same as Fig.~\ref{fig:times-flat} but for
    case 6 - the `SProcess' profile.}
  \label{fig:times-sprocess}
\end{figure}

\begin{table*}
  \caption{Relative computation times.}
  \centering
  \begin{tabular}{l|cccc|cccc|cccc}
    \hline \hline
              & \multicolumn{4}{c|}{Wagoner's Method} & \multicolumn{4}{c|}{Bader-Deuflehard Method} & \multicolumn{4}{c}{Gear's Method}\\ 
              &                & \multicolumn{3}{c|}{Percentage} &  & \multicolumn{3}{c|}{Percentage} &  & \multicolumn{3}{c}{Percentage} \\ 
    Case      & Relative Time & (i)  & (ii) & (iii) & Relative Time & (i) & (ii) & (iii) & Relative Time & (i) & (ii) & (iii) \\ \hline
    1) Flat      & 1    & 20  & 6    & 74    & 0.034         & 1   & 4    &  95   & 0.13           & 11  & 6    & 82    \\
    2) Flat2     & 1    & 19  & 15   & 66    & 0.11          & 2   & 44   &  54   & 0.083          & 5   & 59   & 36    \\
    3) Merger    & 1    & 31  & 10   & 59    & 0.30          & 39  & 1    & 60    & 0.37           & 23  & 9    & 68    \\
    4) Nova      & 1    & 42  & 18   & 40    & 9.7           & 36  & 2    & 62    & 0.80           & 28  & 16   & 56    \\
    5) XRB       & 1    & 24  & 7    & 69    & 1.1           & 38  & 1    & 61    & 0.18           & 19  & 8    & 73    \\
    6) S-process & 1    & 36  & 16   & 48    & 2.5           & 38  & 2    & 60    & 0.58           & 21  & 13   & 66    \\
    \hline \hline
  \end{tabular}
  \label{tab:timefractions}
\end{table*}

\section{Discussion}
\label{sec:discussion}

\subsection{Case 1: Flat}
\label{sec:discuss-flat}

This computationally challenging flat profile represents hydrostatic
helium burning in which helium is consumed to produce ashes largely
consisting of \nuc{52}{Fe}.  As shown in Fig.~\ref{fig:times-flat},
Wagoner's method of integration requires a significant amount of fine
tuning in order to successfully complete the evolution of the network
for this profile. For large values of \YTMIN, the production of
intermediate isotopes is not sufficiently accounted for in order to
successfully complete the evolution. Convergence problems at later
times are a consequence of this loss of accuracy, causing
unrecoverable errors and premature exit from the code. Conversely, for
small values of \YTMIN, convergence problems also arise. In this case,
simplifications required in the semi-implicit method become invalid
for nuclei with small abundances in the presence of rapidly changing
nucleosynthesis flows. To achieve reasonable values of precision for
this profile, integration parameters of \SSC$=1000$ and
\YTMIN$=10^{-8}$ are needed\footnote{Note that the integration
  variables used to compare computation times for each of the other
  cases discussed here could not be used for this profile because they
  caused the computation to exceed our $100\,000$ step limit.}. Using
these parameters, full integration is completed successfully in
2500~s, with solution of equations being responsible for approximately
74\% of that computation time. Recalculation of the Jacobian matrix in
this case requires 20\% of the computation time owing to the large
number of steps required (over $70\,000$).

The Bader-Deuflehard method performs remarkably well for this flat
profile\ \citep[see also][]{TIM99}. In very few time-steps, the method
accurately completes integration over the profile. Furthermore, very
small \YSCALE\ and \EPS\ values can be safely used, resulting in rapid
completion of the nucleosynthesis integration. This test highlights
the power of the Bader-Deuflehard method when the profile is smoothly
varying and the extrapolation method discussed in
Sec.~\ref{sec:th-bader} can be used without additional step-size
limitations. In comparison with Wagoner's method above, we see in
Tab.~\ref{tab:timefractions} that for our chosen set of integration
parameters, the Bader-Deuflehard method successfully completes
integration of the profile rapidly (just 85~seconds and 129 steps,
just 3\% of the time required by Wagoner's method). This small number
of integration steps explains why building the Jacobian matrix
accounts for only 1\% of the total computation time. The method
overhead time, however, is relatively high from computing errors and
step-size estimates. Note that even for extreme values for desired
precision and accuracy, successful integration can be reached in a
reasonable amount of time.

The Gear's method for this profile performs remarkably well at large
values of \YSCALE\ and \EPS, completing evolution within 100
seconds. However, as these parameters increase, the time required for
successful integration increases dramatically. At very small values of
\YSCALE\ and \EPS, evolution is not completed within the $100\,000$
step limit imposed in this work and computation is halted. For
integration parameter values of \YSCALE$=1\times 10^{-10}$ and
\EPS$=1\times 10^{-3}$, the integration is completed in just over 326
seconds for this test case: roughly a factor of three slower than the
Bader-Deuflehard method discussed above. Table~\ref{tab:timefractions}
shows that, for these integration parameters, solution of the
  linear system of equations requires the most time (82\% CPU time),
with method overhead requiring about 6\% of the computational
time. The increased method overhead compared to Wagoner's method is
offset by the total time required for integration, just 326~s and 5000
steps.

\subsection{Case 2: Flat2}
\label{sec:discuss-flat2}

In many nucleosynthesis situations, less extreme conditions are
reached than those considered in the first flat profile. The computation times required by all three methods are shown in
Fig.~\ref{fig:times-flat2}.  Wagoner's method performs considerably
more stably for this profile than for the more extreme flat profile
case. Here, the execution time depends weakly on \YTMIN, and rather
depends quite strongly on \SSC, which effectively forces the code to
take more steps than is necessary for this profile.  Wagoner's method
seems to be sufficient for successful integration in this case. Using
the standard choice of integration variables discussed above,
computation time for this profile is just 64~s with solution
  of the linear system of equations requiring 66\% of that time.

The Bader-Deuflehard method for this profile also performs well, with
most choices of integration parameters resulting in completion times
under 7~s. However, for the most extreme choices of \YSCALE$=10^{-18}$
and \EPS$=10^{-6}$, a considerable increase is shown in
Fig.~\ref{fig:times-flat2} that corresponds to a comparatively large
number of time-steps ($\sim 100$ steps) needed by the accuracy
constraints. This rapid increase highlights the poor performance of
this method under scenarios requiring large numbers of time-steps. For
more reasonable integration parameters, however,
Tab.~\ref{tab:timefractions} shows that this method performs very
well, requiring under one tenth of that needed by Wagoner's method to
complete integration over this profile.

Gear's method also performs very well for this profile, with moderate
dependence of execution time on both \YSCALE\ and \EPS. Even for the
most restrictive case using \YSCALE$=10^{-16}$ and \EPS$=10^{-6}$,
successful completion is still achieved in just 17 seconds. For the more
reasonable parameters discussed before, the profile is successfully
completed in 5 seconds. In this case, the method overhead (i.e.,
computing errors and estimating step-sizes) accounts for a large
fraction of this time at 59\%.

\subsection{Case 3: Merger}
\label{sec:discuss-merger}

\begin{figure}
  \centering
  \includegraphics[width=0.8\hsize]{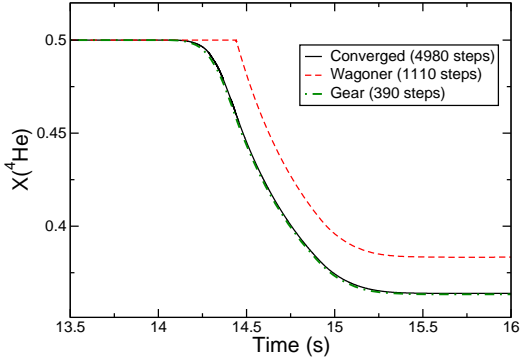}
  \caption{(Colour online) Example of the importance in step error
    checking to reaction network integration. For Wagoner's method
    (red dashed line), the algorithm can step past profile changes and
    therefore incorrectly compute nucleosynthesis activity while
    retaining good conservation of mass (the main tool available for
    estimating accuracy). Gear's method (green dot-dashed line), on
    the other hand, can detect these changes by calculating the
    truncation error of each step, and return for smaller steps in the
    case where convergence fails. The black, solid line represents a
    Wagoner's method calculation using conservative integration
    parameters found to produce converged nucleosynthesis results.}
  \label{fig:overstepping}
\end{figure}

For the merger profile and network test case, results are presented in
Fig.~\ref{fig:times-merger}. It is immediately obvious that Wagoner's
method does not provide reliable results for the smallest values of
\SSC$=10$ chosen in this study, even when integration over the profile
is completed without error. A good example of this effect is shown in
Fig.~\ref{fig:overstepping}, where Wagoner's method using small values
of \SSC\ successfully steps past sudden profile changes, yielding
inaccurate nucleosynthesis compared to our converged results obtained
using a set of carefully chosen parameters. These are obtained by
first picking restrictive integration parameters, and then varying
those parameters to ensure that final abundances do not vary
significantly. The smooth behaviour of the converged results in
Fig.~\ref{fig:overstepping} also indicates that convergence is reached
in this case. This failure to detect sudden changes in the profile
stems from the lack of true error checking in this method. In this
example, a large step is taken from 13.7~s to 14.5~s using an
intermediate point at 14.1~s, where the temperature has not yet
changed, thus leading to unchanged abundances. It is not until the
following step, therefore, that high nuclear interaction activity is
found. At that point, no good mechanism is implemented for
automatically reversing the evolution by multiple time-steps to
re-attempt integration over rapidly varying profiles.  For our set of
safe parameters discussed previously, however, Wagoner's method
does reach accurate abundance values in 115~s and about $27\,000$
steps. Two thirds of the computation time is used for solving
  the linear system of equations, with the other third used for rate
calculation and method overhead.

The Bader-Deuflehard method exhibits interesting properties for this
merger profile. It is immediately apparent that longer computation
times are required to accurately integrate the network over this
profile. This problem arises from the shape of the profile
(Fig.~\ref{fig:merger}). The dynamic, rapidly changing nature of the
merger profile means that the infinite sub-step result of the
Bader-Deuflehard method does not satisfactorily reach convergence for
large time-steps. The shape of the profile changes so quickly that
small steps must be taken to achieve convergence, thus erasing the
benefit of the method. For our safe integration parameter
choices, for example, 790 steps are required. While this represents
half the number of steps required by Gear's method for this profile,
and only a small fraction of the number of steps required by Wagoner's
method, the expense of each step causes total integration time to be
335~s, almost 8 times that required by Gear's method and over twice the
time required by Wagoner's method. Table~\ref{tab:timefractions} shows
that most of this computation time is used to calculate rates ($\sim
40$\%) and solve the systems of equations ($\sim 60$\%).

Gear's method successfully completes evolution with final abundances
in agreement with the converged abundances for all cases except the
most extreme \YSCALE\ and \EPS\ values. Those cases failed because the
step count exceeded the $100\,000$ step limit imposed in this
study. The method's behaviour follows a predictable pattern with more
precise evolution requiring more time. Integration parameter values of
\YSCALE$=1\times 10^{-10}$ and \EPS$=1\times 10^{-3}$ yielded an
integration time of 42~s in about $2800$ steps, representing a speed-up
factor of approximately 3 over Wagoner's method for similar result
accuracy. For this case, the method overhead is larger, but the reward
in total computation time is clear.

\subsection{Case 4: Nova}
\label{sec:discuss-nova}

Consider the Wagoner's method run-times in Fig.~\ref{fig:times-nova}.
Some immediate observations can be made. Small values of \SSC\
(\SSC$=10, 100$) cause nucleosynthesis completion apparently
successfully, but as the small data points indicate, the abundances
obtained from these runs disagree with converged values. Furthermore,
it is difficult to predict which combinations of parameters will
complete the evolution successfully. For example, the combination of
\SSC$\,=1000$ and \YTMIN$\,=10^{-8}$ reaches accurate abundances, but
when \YTMIN$\,=10^{-10}$, incorrect abundances are obtained. Only three of
the 7 runs with \SSC$\,=10^4$ reached accurate abundances. The method
accuracy can be tweaked using other integration parameters, but this
example highlights some of the pitfalls of using network integration
methods that do not include reliable error estimates. For our choice
of safe integration parameters, however, reasonably accurate
results for abundances $X>10^{-8}$ are obtained in just 3~s.

As was the case with the merger profile, the rapidly changing nova
profile limits the step-size available to the Bader-Deuflehard
method. These limited step-sizes result in very long computation times
when using the Bader-Deuflehard method for nova profiles. None of the
integration parameter sets considered here exceeded the $100\,000$ step
limit imposed in our study. For our safe choice of parameters,
successful profile integration was achieved in 34~s and
$1200$ steps. In this case, rate computation was a computationally
expensive part of the calculation, requiring 36\% of the time and
indicating that many sub-steps were required to reach convergence.

Gear's method follows a much more predictable pattern than Wagoner's
method for this profile. It is immediately obvious that most
computations complete the evolution successfully with values that
agree with the converged values. Furthermore, the error checking
routines in Gear's method ensure that the code rarely exits
successfully with inaccurate results outside the range dictated by our
input parameters. This is the immediate benefit of using reliable
truncation error estimates to constrain the method. The computation
time required for these profiles is similar to Wagoner's method, with
run-time for our chosen set of integration parameters taking just
2.8~s. Roughly half of this time is used for solving the
  systems of equations while rate calculation and method overhead
account for 28\% and 16\%, respectively. The total number of
LU-decompositions for this method is similar to that of
Wagoner's method resulting in an integration time improvement of only
20\%, but it is clear that the benefit in this case come in the form
of reliability.

\subsection{Case 5: X-ray burst}
\label{sec:discuss-xrb}

The X-ray burst profile shape is qualitatively similar to that of the
nova profile and thus, one would expect similar behaviour of the
methods. However, the scope of nuclear activity during the profile
dramatically changes the behaviour of Wagoner's method with respect to
the integration parameters shown in Fig.~\ref{fig:times-xrb}. For the
largest values of \YTMIN, Wagoner's method does not successfully
complete integration of the network regardless of the \SSC\ value. In
this case, the underlying reason is the same as that for the high
temperature flat profile (see Fig.~\ref{fig:times-flat} and
Sec.~\ref{sec:discuss-flat}). That is, for large \YTMIN\ values,
nucleosynthesis of less abundant species are not computed with enough
accuracy. Consequently, once they do become abundant, the loss of
accuracy at early times is propagated forward and mass conservation is
broken in the system in an unrecoverable way. However, for more
typical values of \YTMIN, the method is quite robust. Run-times follow
a predictable pattern in which smaller values of \YTMIN\ require more
time for integration. Interestingly, \SSC\ values of $100$ and $1000$
result in almost identical run-times. The reason for this is that for
these cases, the actual time-steps taken for the two values are
smaller than the maximum allowed by \SSC. This is expected behaviour
for this integration method, although it is not easily predictable for
a particular profile. For our choices of \SSC$=1000$ and
\YTMIN$=10^{-12}$, network integration is completed in 1300~s and
$45\,000$ steps, highlighting the challenging nature of computing
nucleosynthesis in X-ray bursts. Most of this time (69\%) is required
for solving the systems of equations, with rate calculations
at each step taken 24\% of the computation time.

The Bader-Deuflehard method behaves similarly for X-ray burst
nucleosynthesis as for novae. For large \EPS\ values, convergence is
achieved, but the time required is similar to that of Wagoner's
method. These run-times are explained similarly as for nova and merger
profiles. The allowed time-step becomes limited by the profile rather
than by the method itself. For small time-steps, the advantages of the
method are outweighed by the large number of LU-decompositions
necessary. 

In direct contrast with the Bader-Deuflehard method, Gear's method run
times are remarkably fast for the X-ray burst model. Once again, the
behaviour of run time as a function of the integration parameters
follows a predictable pattern, but with small \EPS\ and \YTMIN\
combinations not fully completing integration within our maximum step
limit. As Tab.~\ref{tab:timefractions} shows, integration over this
profile takes about 240~s ($\sim 6100$ steps) with reasonable
parameters: over 5 times faster than Wagoner's method and the
Bader-Deuflehard method. With this choice of parameters, we find that
the integration accuracy achieved is good down to the $X=10^{-12}$
level. The same pattern is found as before in which the method
overhead is larger than it is for the other methods investigated in
this work, but is offset by the robust and efficient performance
overall.

\subsection{Case 5: s-Process}
\label{sec:discuss-sproxess}

Finally, the s-process profile shown in Fig.~\ref{fig:sprocess} is
considered, with integration times presented in
Fig.~\ref{fig:times-sprocess}. Wagoner's method, once again,
unsuccessfully complete nucleosynthesis integration over this profile
for large values of \YTMIN. While in this case there are no sharp
profile changes, there is a sudden nucleosynthesis activity change
when the \Nepa\ reactions become active towards the end of the
profile. Unless the evolution of nuclei is carefully accounted for at
low abundance levels, they are not guaranteed to fulfil mass
conservation requirements once they reach appreciable levels. However,
for most higher values of \YTMIN, Wagoner's method performs rather well
for this s-process nucleosynthesis profile. For our selection of
safe integration parameters discussed before, for example, the
profile is integrated over in 36~s and $5500$ steps. Rate calculation
times become important for these small computation times, totalling
36\% of the total computation time.

The Bader-Deuflehard method varies in performance for this test
case. While, for large values of \YSCALE and \EPS, integration times
are on par with those from Wagoner's method, for the smallest values
it can perform rather poorly. In this case, it is the nature of the
nucleosynthesis itself that limits the optimal step-sizes required to
achieve convergence in the routine. For values of \EPS$\,=10^{-6}$,
the Bader-Deuflehard method requires up to a factor of 10 times more
computational time than Wagoner's methods to complete nucleosynthesis
integration. For our representative set of integration parameters,
this method is about 2.5 times slower than Wagoner's method.
 
Gear's method in this case follows much the same behaviour as for
other cases. Once again, the most extreme cases of small \YSCALE\ and
\EPS\ require many small time-steps that exceed the limits adopted in
this work. For more reasonable values, however, computation times are
comparable with Wagoner's method, although a larger fraction of this
time is required for solving the linear systems of
  equations. For our representative integration parameters, the
Gear's method requires 20~s in 1800 steps, highlighting the relative
cost of each step in the method in comparison to Wagoner's method,
although still performing the integration twice as fast.

\section{Conclusions}
\label{sec:conclusions}

Integration of nuclear reaction networks has not received much
modernisation since the first stellar modelling codes were developed
over 4 decades ago. Now that computational power is becoming more
available, it is possible to compute models in a relatively short
period of time, thus opening avenues for detailed studies of the
effects of varying model parameters. 
It is worth re-visiting the question of integration method efficiency
to implement faster, more reliable nucleosynthesis solvers.

In this work, we tested the performance of two integration methods:
the Bader-Deuflehard method and Gear's backward differentiation method
in comparison to the traditionally used Wagoner's method. To fully
investigate the behaviour of these integration methods, a suite of
test profiles was used. This suite included 2 flat profiles, and
profiles representing temperatures and densities in white dwarf
mergers (helium burning), nova explosions (high temperature hydrogen
burning), thermonuclear runaway, in X-ray bursts (the rp-process), and
core helium burning in massive stars (He-burning and s-process
nucleosynthesis).

In agreement with the findings of \cite{TIM99}, we found that although
Wagoner's method can sometimes be a fast way to integrate nuclear
reaction networks, the lack of error estimates means that accuracy of
the end results can be hard to predict. For more challenging cases
such as nova or X-ray burst nucleosynthesis, great care must be taken
to ensure accurate results. 
Even small variations in integration parameters can sometimes yield
wildly varying results.
The Bader-Deuflehard and Gear's methods, on the other hand, rarely
report inaccurate results at the precision desired through setting
integration parameters. As reported in \cite{TIM99}, the
Bader-Deuflehard method was found to be very powerful for flat
profiles in which the step-sizes were only limited by the accuracy of
the method. However, when dramatically varying temperature-density
profiles were introduced, this was not necessarily the case,
  at least in a post-processing framework. Once small step-sizes are
required to follow a sharp profile, the cost of each individual step
in this method increases the computation time considerably. For simple
profiles, or for use within a hydrodynamics code in which step-size
can be safely increased to large values, this method is very robust
and powerful.

Finally, Gear's method was found to be very robust, only
encountering difficulties for tight constraints on desired precision
and accuracy. For all profiles in which the step-size was limited by a
rapidly changing temperature and density, this method was found to
out-perform both Wagoner's and the Bader-Deuflehard
methods. Furthermore, the speed of computation followed a clear trend
with integration parameters, making it simple to use and easy to
predict. The main disadvantage of this method is the difficulty in
implementing it, since past behaviour must be stored in such a way to
account for failed steps or changes of integration order. For more
advanced applications, however, this method should be considered a
powerful method for integrating nuclear reaction networks.

Summarising, we show that both the Bader-Deuflehard and Gear's methods
exhibit dramatic improvements in both speed and accuracy over 
Wagoner's method used by many codes. Moreover, in applications for which
the environment is expected to change rapidly, we found Gear's
method to be most robust and, thus, recommend its use in stellar
codes.

\begin{acknowledgements}
  This work has been partially supported by the Spanish grants
  AYA2010-15685 and by the ESF EUROCORES Program EuroGENESIS through
  the MICINN grant EUI2009-04167. We also thank the referee, Frank
  Timmes, for invaluable suggestions that helped to improve the
  manuscript.
\end{acknowledgements}

\bibliographystyle{aa}
\bibliography{bib}

\end{document}